%% file: No-lossy10-extension.tex
\documentclass[11pt]{article}
\usepackage{amstext,amssymb,amsmath,amsbsy}
\usepackage{hyperref}
\usepackage{amscd}
\usepackage{amsfonts}
\usepackage{indentfirst}
\usepackage{verbatim}
\usepackage{amsmath}
\usepackage{amsthm}
\usepackage{enumerate}
\usepackage{graphicx}
\usepackage[OT1]{fontenc}
\usepackage[latin1]{inputenc}
\usepackage[english]{babel}
\usepackage{amssymb}
\input{macros}

\numberwithin{equation}{section}

\title{Approximate cloaking for the Helmholtz equation via transformation optics and consequences for perfect cloaking}
\author{Hoai-Minh Nguyen \footnote{Courant Institute, NYU, 251 Mercer Street, New York, NY 10012,  USA, hoaiminh@cims.nyu.edu}}

\date{April 07, 2011}

\begin{document}
\maketitle

\begin{abstract}
In this paper,  we study approximate cloaking of active devices for the Helmholtz equation in the whole space of dimension $2$ or $3$ using the scheme introduced by Kohn, Shen, Vogelius, and Weinstein in \cite{KohnShenVogeliusWeinstein}. More precisely, we assess the degree of invisibility, determine the limit of the field inside the cloaked and cloaking regions, and show that the scheme is unstable with respect to the material parameters. As a consequence, we obtain some feasible properties of ``perfect" cloaking.
\end{abstract}

%\tableofcontents

\section{Introduction}

Cloaking via change of variables was introduced by Greenleaf, Lassas, Uhlmann \cite{GreenleafLassasUhlmann} for electrical impedance tomography, Pendry, Schurig, and Smith \cite{PendrySchurigSmith} for the Maxwell system, and Leonhardt \cite{Leonhardt} in the geometric optics setting. They used a singular change of variables which blows up a point into a cloaked region. This singular structure implies not only difficulties in practice, but also in analysis. Some approaches \cite{GreenleafKurylevLassasUhlmann07}, \cite{WederRigorous}, \cite{WederRigousTimeDomain} are proposed to tackle this problem mathematically  based on the notion of ``weak" solution. To avoid using the singular structure, regularized schemes have been proposed in \cite{Schurig06}, \cite{Kildishev07}, \cite{YanRuanQiu07}, \cite{RuanYanNeffQiu}, \cite{KohnShenVogeliusWeinstein}, and \cite{GreenleafKurylevLassasUhlmann07SHS}. The reader can find more information and references related to cloaking in the works mentioned above or in the reviews \cite{GreenleafKurylevLassasUhlmann} and \cite{Wood}.

\medskip
In this paper, we study approximate cloaking of active devices for the Helmholtz equation in the whole space of dimension $2$ or $3$ for the scheme introduced by Kohn, Shen, Vogelius, and Weinstein in \cite{KohnShenVogeliusWeinstein}, where they used a transformation which blows up a small region instead of a point into the cloaked region.
%we use the scheme introduced by Kohn, Shen, Vogelius, and Weinstein in \cite{KohnShenVogeliusWeinstein}, where they used a transformation which blows up a small ball instead of a ball into the cloaked region, to study approximate cloaking of active devices for the Helmholtz equation in the whole space of dimension $2$ or $3$. We assume that the cloaked region as well as the material parameters inside are fixed.
We assess the degree of invisibility, determine the limit of the field inside the cloaked and cloaking regions, and show that the scheme is unstable with respect to the material parameters. As a consequence, we obtain some feasible properties of ``perfect" cloaking. More precisely, let $\eps$ denote the parameter of regularization i.e.  the ratio between the diameter of the region blown up to the cloaked region $D$ and the diameter of the cloaked region:

\medskip
$a)$ In the $3d$ non-resonant case i.e. when the frequency is not an eigenvalue of the Neumann problem in $D$, we show that the difference between the field and the ``push-forward" of the solution of the Helmholtz equation in free space is of order $\eps$ in any bounded region away from $D$ (Theorem~\ref{thm1-3d}). Concerning the limiting behavior of the approximate field in $D$, we prove that it converges and the limit is the solution of the corresponding Neumann problem in $D$ (Theorem~\ref{thm1-3d} and Definition~\ref{defCl3}).

\medskip
$b)$ In the $3d$ resonant case i.e. when the frequency is an eigenvalue of the Neumann problem in $D$, under the assumption that the source inside the cloaked region is orthogonal, with respect to the $L^2$-scalar product in $D$, to all Neumann eigenfunctions \footnote{This condition is necessary and sufficient to have a solution of the Neumann problem. This is a consequence of the Fredholm alternative theory.} (this holds if the cloaked region is passive), we also prove that the difference between the field and the ``push-forward" of the solution of the Helmholtz equation in free space is of order $\eps$ in any bounded region away from $D$, the approximate field converges in $D$ and the limit is uniquely determined. However, this limit is not only a solution of the  Neumann problem in $D$, but also depends on the value of the solution of the Helmholtz equation in free space at the point where the maps are blown up in a quite involved manner (Theorem~\ref{thm1-3d} and Definition~\ref{defCl3}).

\medskip
$c)$ In the $3d$ resonant case, without the assumption on the orthogonality of the source mentioned in statement $b)$, the energy of the field inside $D$ explodes as $\eps \to 0$; moreover, cloaking can be not achieved (Proposition~\ref{proInfinite}).

\medskip
$d)$ In the $2d$-non resonant case (see the definition in Definition~\ref{defNonRes}), we show that the field converges to the ``push-forward" of the solution of the Helmholtz equation in free space in any bounded set away from $D$  with a rate $1/ |\ln \eps|$ and the limiting behavior of the field in $D$ exhibits a non-local structure. Therefore, the limit is generally not the solution of the  Neumann problem in $D$ (Theorem~\ref{thm1-2d} and Definition~\ref{defCl2}).

\medskip
$e)$ In the $2d$-resonant case (see the definition in Definition~\ref{defNonRes}), we prove that the energy inside the cloaked region can go to infinity and cloaking can be not achieved (Proposition~\ref{proInfinite-2d}).

\medskip $f)$ Concerning the stability with respect to the material parameters inside the cloaked region without a source, we show that cloaking is unstable with respect to these parameters. Roughly speaking, there exist some fixed parameters such that for each $\eps$, one can perturb these parameters by an amount of order $\eps$ in $3d$ and $1/ |\ln \eps|$ in $2d$ so that the degree of visibility is of order 1 (Proposition~\ref{proinstability}).

\begin{remark} Property $f)$ does not contradict Properties $a)$ and $d)$ since Property $f)$ is only stated under a condition on the amount of the perturbation which does not hold when the material parameters are fixed.
\end{remark}

%\begin{remark} We emphasize here that in $2d$ our result is not a consequence of the zero frequency one since we first fix the frequency, then let $\eps$ go to $0$.
%\end{remark}

\begin{remark} A similar conclusion as in statement $f)$ for  $2d$ bounded domains is previously observed by Kohn et al. \cite{KohnOnofreiVogeliusWeinstein}.
\end{remark}

Our results in the $3d$ non-resonant case are compatible with what has been mentioned in the literature: cloaking is achieved, the energy of the field is finite in the cloaked region, and the limit of the field inside the cloaked region is the solution of the Neumann problem see e.g. \cite{GreenleafKurylevLassasUhlmann07}, \cite{WederRigousTimeDomain}. However, our results  in the $3d$ resonant case are quite different. In the setting in \cite{GreenleafKurylevLassasUhlmann07}, the limit field could be any Neumann eigenfunction in the cloaked region for a passive device. In \cite{WederRigousTimeDomain}, the author asserted that the field inside and outside the cloaked region are completely decoupled from each other. It is discussed in the literature that: cloaking for the Helmholtz equation is achieved and the energy of the field inside the cloaked region is finite (see e.g. \cite{GreenleafKurylevLassasUhlmann07}, \cite{Cummer08}, \cite{WederRigorous}, \cite{WederRigousTimeDomain}).
The $2d$ perfect cloaking has not been rigorously studied as extensively as in the $3d$ one and it is often argued that the field in the cloaked region is a solution of the Neumann problem.

\medskip
We recall that the weak solution considered in \cite{GreenleafKurylevLassasUhlmann07} is only discussed in $3d$. The degree of invisibility in the approximate cloaking problem is more widely understood when one uses an appropriate lossy-layer. In this case, the same estimates as above hold independent of the material parameters in $D$ (see \cite{KohnOnofreiVogeliusWeinstein}, \cite{NguyenHelmholtz}, \cite{NguyenVogelius2}) and there are explicit frequency dependent estimates   which are valid for all frequency (see \cite{NguyenVogelius2}). The zero frequency case is less complicated. This is studied in \cite{KohnShenVogeliusWeinstein} (see also \cite{NguyenVogelius}) where no lossy-layer is used and better estimates are obtained. Without a lossy-layer, the degree of invisibility in $2d$ is discussed in \cite{RuanYanNeffQiu} when the material parameters inside the cloaked region are isotropic and homogeneous, and the approximate cloaking is confirmed for the $3d$ non-resonant case in \cite{GreenleafKurylevLassasUhlmann08Approximate} (without an estimate of the degree of invisibility).
Recently, Greenleaf, Kurulev, Lassas, and Uhlmann \cite{GreenleafKurylevLassasUhlmann09CloakingShielding} observed that cloaking without shielding is possible (compare with our results in the $3d$ resonant case).

\medskip
Let us describe the problem more precisely. To illustrate the idea, let us suppose that the cloaking region is the annular $\{1 < |x| < 2 \}$ and the cloaked region is the unit ball $B$ centered at the origin of $\mR^d$ ($d=2,3$). Using the scheme in \cite{KohnShenVogeliusWeinstein}, the parameters in the cloaking region are given by
\begin{equation*}
a_c, \sigma_c = {F_\eps}_{*}I, {F_\eps}_{*}1,
\end{equation*}
where $F_\eps$ is the maps which blows up the ball $B_\eps$ into $B_1$ given by
\begin{equation}\label{defFeps}
F_\eps = \left\{\begin{array}{cl} x & \mbox{ if } x \in \mR^d \setminus B_2, \\[12pt]
\dsp \Big(\frac{2 - 2\eps}{2 - \eps} + \frac{|x|}{2 - \eps} \Big) \frac{x}{|x|} & \mbox{ if } x \in B_2 \setminus B_{\eps}, \\[12pt]
\dsp \frac{x}{\eps}  & \mbox{ if } x \in B_{\eps}.
\end{array}\right.
\end{equation}
Hereafter we use the standard notation
\begin{equation}\label{defF*}
F_*A(y) = \frac{D F (x) A(x) DF^T(x)}{ \det DF(x)} \quad \mbox{and} \quad  F_*\Sigma(y) = \frac{\Sigma(x)}{ \det DF(x)}, \quad \mbox{ where } x = F^{-1}(y),
\end{equation}
for any matrix-valued function $A$ and any complex function $\Sigma$ and we denote $D_r$ the set $\{r x; x \in D \}$ for any open bounded set $D$ of $\mR^d$ and for any $r >0$.

\medskip
Let $a$ be a uniformly elliptic matrix-valued function defined in $B_1$, $\sigma$ be a complex function defined in $B_1$, $f\in L^2(\mR^d)$ such that $0 \le \ess \inf \Im (\sigma) \le \ess \sup \Im (\sigma)  < + \infty$, $0 < \ess \inf \Re (\sigma) < \ess \sup \Re (\sigma) < + \infty$, $\supp f \subset (B_4 \setminus B_3) \cup \overline{B_1}$. For $k>0$, let  $u, u_c \in H^1_{\loc}(\mR^d)$ be the unique outgoing solutions of the equations
\begin{equation}\label{eq-u}
\Delta u + k^2 u = f \chi_{\{|x| \ge 2 \}} \quad \mbox{ in } \mR^d,
\end{equation}
and
\begin{equation}\label{eq-uc}
\dive(A_c \nabla u_c) + k^2 \Sigma_c  u_c = f \quad \mbox{ in } \mR^d,
\end{equation}
where $\chi_{\Omega}$ denotes the characteristic function of the set $\Omega$ for any $\Omega \subset \mR^d$, and
\begin{equation}\label{defASigmac}
A_c, \Sigma_c = \left\{\begin{array}{cl}
I,1 & \mbox{ in } \mR^d \setminus B_2, \\[12pt]
a_c = {F_\eps}_{*}I, \sigma_c = {F_\eps}_{*}1 & \mbox{ in } B_2 \setminus B_1, \\[12pt]
a, \sigma & \mbox{ in } B_1.
\end{array} \right.
\end{equation}
In the following, whenever we talk about outgoing solutions to a Helmholtz problem at frequency $k$, we mean solutions that satisfy
\begin{equation*}
\dsp \frac{\partial v}{\partial r} - i k v = o(r^{-\frac{d-1}{2}}) \quad \mbox{ as } r = |x| \lr \infty,
\end{equation*}
where $d=2$ or 3 is the dimension of space.

\begin{remark}
Physically $a$, $\sigma$ are the material parameters in the cloaked region $B_1$; $f$ is a given source  outside the cloaking region, $u$ describes the field corresponding to the source existing outside the cloaked and cloaking regions in free space, and $u_c$ describes the field due to the source $f$ in the presence of the cloaking device and the cloaked object.
\end{remark}

The effectiveness of the approximate cloak and the limit of $u_c$ as $\eps \to 0$ are described in Theorems~\ref{thm1-3d} and \ref{thm1-2d} below. The proofs will be presented in Sections~\ref{Section-3d} and \ref{Section-2d}.

\begin{theorem}\label{thm1-3d} Let $d=3$, $k > 0$, and $0< \eps < 1$. Define \footnote{In the following, $\eta$ denotes the unit normal vector on $\partial D$ directed to the complement of $D$ for any smooth bounded open subset $D$ of $\mR^d$.}
\begin{equation}\label{defMB1}
M: = \Big\{\psi \in H^1(B_1); \; \dive (a \nabla \psi) + k^2 \sigma \psi = 0 \mbox{ in } B_1 \mbox{ and } a \nabla \psi \cdot \eta = 0 \mbox{ on } \partial B_1 \Big\}.
\end{equation}
In the resonant case i.e. when $M \not = \{ 0 \}$, assume in addition that $\dsp \int_{B_1} f \bar \be = 0$ for any $\be \in M$ \footnote{This condition  is necessary and sufficient to have a solution $v \in H^1(B_1)$ of the system $\dive (a \nabla v) + k^2 \sigma v = f$ in $B_1$ and $a \nabla v \cdot \eta = 0$ on $\partial B_1$ by the Fredholm alternative theory see e.g. \cite{BrAnalyse} or \cite{LaxFunct}.}. Then for all $K \subset \subset \mR^3 \setminus \overline B_1$,
\begin{equation*}
\| u_c - u \circ F_0^{-1} \|_{H^1(K)}  \le C \eps \| f \|_{L^2},
\end{equation*}
for some positive constant $C$ depending on $k$, $K$, $a$, and $\sigma$, but independent of $\eps$, $f$. We also have that $u_c \rightharpoonup Cl_3(f)$ weakly in $H^1(B_1)$. As a consequence
\begin{equation*}
\lim_{\eps \to 0} \| u_c - Cl_3(f)\|_{L^2(B_1)} = 0.
\end{equation*}
\end{theorem}

Here and in the following, $F_0(x) := \lim_{\eps \to } F_\eps(x)$ for $x \in \mR^d \setminus \{ 0\}$  and for  $d=2,3$.

\medskip

The operator $Cl_3$ is defined in the following

\begin{definition}\label{defCl3} Let $d=3$.  We define $Cl_3$ as follows:
\begin{enumerate}
\item[i)] The non-resonant case: $M = \{ 0 \}$. Define $Cl_3(f) =v$ where $v \in H^1(B_1)$ is the unique solution of the system
\begin{equation*}\left\{
\begin{array}{ll}
\dive(a \nabla  v) + k^2 \sigma v = f & \mbox{ in } B_1, \\[12pt]
\dsp a \nabla v \cdot \eta \Big|_{\mathrm{int}} = 0 & \mbox{ on } \partial B_1.
\end{array}\right.
\end{equation*}

\item[ii)] The resonant case: $M \not = \{ 0 \}$. Assume in addition that $\dsp \int_{B_1} f \bar \be = 0$ for any $\be \in M$. Consider the triple of functions
$(v_{\mathrm{ext}}, v_{\mathrm{int}}, w) \in W^1(\mR^3 \setminus \overline B_1) \times H^1(B_1) \times M^{\perp}$ which is the unique solution of the systems
\begin{equation*}\left\{
\begin{array}{ll}
\Delta v_{\mathrm{ext}}  = 0 & \mbox{ in } \mR^3 \setminus \overline B_1, \\[12pt]
\dive(a \nabla  v_{\mathrm{int}}) + k^2 \sigma v_{\mathrm{int}} = f & \mbox{ in } B_1, \\[12pt]
\dive(a \nabla  w) + k^2 \sigma w = 0 & \mbox{ in } B_1,
\end{array}\right.  \mbox{ and }
\left\{
\begin{array}{ll}
\dsp v_{\mathrm{ext}} - v_{\mathrm{int}} = - u(0) & \mbox{ on } \partial B_1, \\[12pt]
\dsp a \nabla v_{\mathrm{int}} \cdot \eta  = 0 & \mbox{ on } \partial B_1, \\[12pt]
\dsp \frac{\partial v_{\mathrm{ext}}}{\partial \eta}   = a \nabla w \cdot \eta  & \mbox{ on } \partial B_1.
\end{array}\right.
\end{equation*}
Define
\begin{equation*}
Cl_3(f) = v_{\mathrm{int}}.
\end{equation*}
\end{enumerate}
\end{definition}

Hereafter for a connected smooth open region $U$ of $\mR^3$ with a bounded complement (this includes $U = \mR^3$), $W^1(U)$ is defined as follows \footnote{The space $W^1$, which is defined here in $3d$ and later in $2d$ (see Definition~\ref{defCl2}), has been used in the study of the Laplace equation in an unbounded domain e.g. \cite{Nedelec}.}
\begin{equation}\label{defW13d}
W^{1}(U) = \Big\{ \psi \in L^1_{loc}(U)~;~~ \frac{\psi(x)}{\sqrt{1 + |x|^2}}  \in L^2(U) \mbox{ and } \nabla \psi \in L^2(U) \Big\}.
\end{equation}
On the boundary $\partial D$ of any bounded open set $D \subset \mR^3$, $\phi \Big|_{\mathrm{ext}}$ and $\phi\Big|_{\mathrm{int}}$ denote the trace of $\phi$ from the outside and the inside. For any closed subspace $M$ of $H^1(D)$, we also denote the space $M^{\perp}$ by
\begin{equation}\label{defMorth}
M^{\perp} = \Big\{ \psi \in H^1(D); \; \int_{D} \Big( \nabla \psi \nabla \bar \phi + \psi \bar \phi \Big) \, dx = 0 \quad \forall \, \phi \in M \Big\}.
\end{equation}

\begin{remark} The uniqueness of the triple $(v_{\mathrm{ext}}, v_{\mathrm{int}}, w)$ is established in Lemma~\ref{lemUniqueness-3d} (Section~\ref{Section-3d}). The existence of $(v_{\mathrm{ext}}, v_{\mathrm{int}}, w)$ will follow from the proof of Theorem~\ref{thm1-3d}.
\end{remark}

The following definition will be used in Theorem~\ref{thm1-2d} concerning the $2d$ setting.

\begin{definition}\label{defNonRes} \label{defCl2} In the $2d$ setting, the system is non-resonant if the following problem
\begin{equation}\label{uniqueness2d}
\left\{\begin{array}{ll}
\Delta w= 0 & \mbox{in } \mR^2 \setminus B_1, \\[12pt]
\dive (a \nabla w) + k^2 \sigma w = 0 & \mbox{in } B_1, \\[12pt]
\dsp \frac{\partial w}{\partial \eta}\Big|_{\mathrm{ext}}  =   a \nabla w \cdot \eta  \Big|_{\mathrm{int}} & \mbox{on } \partial B_1.
\end{array}\right.
\end{equation}
only has the zero solution in $W^1(\mR^2)$. Otherwise, the system is resonant. In the non-resonant case, we define $CL_2(f) = v$ the unique solution in $W^1(\mR^2)$ of the system \footnote{The existence of $v$ is a consequence of the Fredholm alternative theory (see also part $ii)$ of Lemma~\ref{lemd=2})}
\begin{equation*}
\left\{\begin{array}{ll}
\Delta v= 0 & \mbox{in } \mR^2 \setminus B_1, \\[12pt]
\dive (a \nabla v) + k^2 \sigma v = f & \mbox{in } B_1, \\[12pt]
\dsp \frac{\partial v}{\partial \eta}\Big|_{\mathrm{ext}}  =   a \nabla v \cdot \eta  \Big|_{\mathrm{int}} & \mbox{on } \partial B_1.
\end{array}\right.
\end{equation*}
\end{definition}

Hereafter for a connected smooth open region $U$ of $\mR^2$ with a bounded complement (this include $U = \mR^2$), $W^1(U)$ is defined as follows
\begin{equation}\label{defW12d}
W^{1}(U) = \Big\{ \psi \in L^1_{loc}(U): \frac{\psi(x)}{\ln(2 + |x|) \sqrt{1 + |x|^2}}  \in L^2(U) \mbox{ and } \nabla \psi \in L^2(U) \Big\}.
\end{equation}

\begin{theorem}\label{thm1-2d} Let $d=2$, $k > 0$, and $0 < \eps < 1$. Assume that the system is non-resonant. Then for all $K \subset \subset \mR^2 \setminus \overline B_1$,
\begin{equation*}
\| u_c - u \circ F_0^{-1} \|_{H^1(K)}  \le \frac{C}{|\ln \eps|} \| f \|_{L^2},
\end{equation*}
for some positive constant $C$ depending on $K$, $k$, $a$, and $\sigma$, but independent of $\eps$ and $f$. We also have that  $u_c \rightharpoonup Cl_2(f)$ weakly in $H^1(B_1)$. As a consequence,
\begin{equation*}
\lim_{\eps \to 0} \| u_c - Cl_2(f)\|_{L^2(B_1)} = 0.
\end{equation*}
\end{theorem}

\begin{remark}From Theorem~\ref{thm1-2d}, the limit of the field in $D$ is $0$ when the cloaked region is passive in the $2d$ non-resonant case.
\end{remark}

We also show in the following proposition that if $k$ is small enough (the smallness condition depends only on the bounds of $a$ and $\sigma$) then the system is non-resonant. Proposition~\ref{proNonRes} will be proved in Section~\ref{Section-2d}.

\begin{proposition}\label{proNonRes} Let $0< c_1 < c_2 < 0$. Assume that $c_1 |\xi|^2 \le \langle a \xi,  \xi \rangle \le c_2 |\xi|^2$, $0 \le \Im (\sigma) \le c_2$, $c_1 < \Re (\sigma) < c_2$ in $B_1$. There exists $k_0 > 0$, depending only on $c_1$ and $c_2$, such that if $k < k_0$, then the system is non-resonant.
\end{proposition}

\medskip

The following proposition, which will be proved in Section~\ref{SectionInfinite}, establishes the results mentioned in statement $c)$.

\begin{proposition} \label{proInfinite} Let $d=3$ and $k > 0$. Assume that $M \not = 0$ and fix an element $\be \in M$ such that $\| \be \|_{L^2(B_1)} =1$. Let $u_c$ be the solution of \eqref{eq-uc} with $f = 0$ in $\mR^3 \setminus B_1$  and $f = \be$ in $B_1$.
\begin{enumerate}
\item[i)] We have
\begin{equation*}
\liminf_{\eps \to 0}  \eps \| u_c\|_{H^1(B_1)} > 0.
\end{equation*}

\item[ii)] Assume in addition that $\be$ is radial, $a$ and $\sigma$ are isotropic and homogeneous in $B_1$ i.e.  $a = \lambda_1 I$ and $\sigma= \lambda_2$ for some positive constants $\lambda_1$ and $\lambda_2$. Then
\begin{equation*}
\liminf_{\eps \to 0} \| u_c\|_{L^2(B_4 \setminus B_2)} > 0.
\end{equation*}
\end{enumerate}
\end{proposition}

Concerning the $2d$-resonant case, we have the following proposition which establishes the results in statement e) and is proved in Section~\ref{SectionInfinite-2d}.

\begin{proposition} \label{proInfinite-2d} Let $d=2$ and $k > 0$. Assume that the system is resonant. Define
\begin{equation*}
N = \Big\{ \psi \in W^1(\mR^2); \, \psi \mbox{ satisfies } \eqref{uniqueness2d}\Big\}.
\end{equation*}
Fix an element $\be \in N$ such that $\| \be \|_{L^2(B_1)} =1$. Let $u_c \in H^1_{\loc}(\mR^2)$ be the unique outgoing solution of \eqref{eq-uc} with $f = 0$ in $\mR^2 \setminus B_1$  and $f = \be$ in $B_1$.
\begin{enumerate}
\item[i)] We have
\begin{equation*}
\liminf_{\eps \to 0}  \| u_c\|_{H^1(B_1)} = + \infty.
\end{equation*}

\item[ii)] Assume in addition that $\be$ is radial, $a$ and $\sigma$ are isotropic and homogeneous in $B_1$ i.e.  $a = \lambda_1 I$ and $\sigma= \lambda_2$ for some positive constants $\lambda_1$ and $\lambda_2$. Then
\begin{equation*}
\liminf_{\eps \to 0} \| u_c\|_{L^2(B_4 \setminus B_2)} > 0.
\end{equation*}
\end{enumerate}
\end{proposition}

Concerning the instability of the approximate cloaking with respect to the material parameters inside the cloaked region, we establish the following result which is proved in Section~\ref{SectionInstability}:

\begin{proposition}\label{proinstability}  Let $d=2,3$, $k>0$, and $\eta \in \mR^d$ with $|\eta| =1$. There exist a positive number $\sigma_0>0$ and a family of positive numbers $(\sigma_\eps)$ such that
\begin{equation}\label{condIns}\left\{
\begin{array}{cl}
0 < \liminf_{\eps \to 0} \eps^{-1} |\sigma_\eps - \sigma_0| \le \limsup_{\eps \to 0} \eps^{-1} |\sigma_\eps - \sigma_0| < + \infty & \mbox{ if } d= 3, \\[12pt]
0 < \liminf_{\eps \to 0} |\ln \eps| |\sigma_\eps - \sigma_0| \le \limsup_{\eps \to 0} |\ln \eps| |\sigma_\eps - \sigma_0| < + \infty & \mbox{ if } d= 2,
\end{array}\right.
\end{equation}
and
\begin{equation*}
\liminf_{\eps \to 0} \| u_{c, s} \|_{L^2(B_4 \setminus B_2)} > 0.
\end{equation*}
Here $u_{c, s} \in H^1_{\loc}(\mR^d)$ is such that $u_{c, s}$ satisfies the outgoing condition and $u_{c}(x):  = u_{c, s}(x) + e^{i k \eta \cdot x}$ is a solution of the equation
\begin{equation*}
\dive(A_c \nabla u_c) + k^2 \Sigma_c u_c = 0.
\end{equation*}
Here $(A_c, \Sigma_c)$ is defined in \eqref{defASigmac} with $a = I$ and $\sigma = \sigma_\eps$.
\end{proposition}

\begin{remark} We recall that when the parameters $a$ and $\sigma$ are fixed and the cloaked region is passive, cloaking is achieved in the $3d$ and $2d$ non-resonant cases (see Theorems~\ref{thm1-3d} and \ref{thm1-2d}). Nevertheless Proposition~\ref{proinstability} does not contradict this fact since it holds under condition \eqref{condIns} which is invalid for fixed $a_\eps$ and $\sigma_\eps$.
\end{remark}

%\begin{remark} The definition of the space $W^1(U)$ in $2d$ and $3d$ cases is taken from \cite[page 59]{Nedelec}.
%\end{remark}

Our approach to Theorems~\ref{thm1-3d} and \ref{thm1-2d} is based on the study of the effect of a small inclusion. The study of approximate cloaking based on the effect of a small inclusion were discussed in \cite{KohnShenVogeliusWeinstein}, \cite{Liu}, \cite{NguyenVogelius}, \cite{KohnOnofreiVogeliusWeinstein}, \cite{NguyenHelmholtz}, \cite{NguyenVogelius2}. It is well-known that when material parameters inside a small inclusion are, roughly speaking, bounded from below and above by positive constants, the effect of the small inclusion is small (see e.g. \cite{VogeliusVolkov}). Without this assumption, the effect of the inclusion is not small (see e.g. \cite{KohnOnofreiVogeliusWeinstein}, \cite{NguyenHelmholtz}) unless there is an appropriate lossy-layer (see \cite{KohnOnofreiVogeliusWeinstein}, \cite{NguyenHelmholtz}, \cite{NguyenVogelius2}). In our setting, the boundedness assumption is violated and no lossy-layer is used. Nevertheless, the effect of the small inclusion is still small (in the non-resonant case) due to the special structure induced from \eqref{defF*}. The starting point of our approach relies on the following well-known fact:

\begin{proposition}\label{fundPro} Let $d \ge 2$, $k>0$, $A$ be a bounded matrix-valued function and $\Sigma$ be a bounded complex function defined on $\mR^d$, $h \in L^2(\mR^d)$, and $F: \mR^d \lr \mR^d$ be a Lipschitz, surjective, and invertible with $F(x) = x$ on $\mR^d \setminus B_2$, and  $\det D F > c$ a.e. $x \in \mR^d$, for some positive constant $c$. Then $u \in H^1_{\loc}(\mR^d)$ is a solution of
\begin{equation*}
\dive (A \nabla u) + k^2 \Sigma u = h \quad \mbox{ in } \mR^d
\end{equation*}
if and only if $v: = u \circ F^{-1} \in H^1_{\loc}(\mR^d)$ is a solution of
\begin{equation*}
\dive (F_*A \, \nabla v) + k^2 F_*\Sigma \,  v = F_*h \quad \mbox{ in } \mR^d,
\end{equation*}
where $F_*A$ and $F_*\Sigma$ are defined in \eqref{defASigmac}, and $F_* h$ is similarly defined as $F_*\Sigma$. Moreover $u = v$ outside $B_2$.
\end{proposition}

Finally we want to mention that the approximate cloaking for the wave equation has been recently studied in \cite{NguyenVogeliuswaves} where an appropriate lossy-layer is used.

\medskip
The paper is organized as follows. Section~\ref{Section-3d} is devoted to the proof of Theorem~\ref{thm1-3d}. In Section~\ref{Section-2d} we prove Theorem~\ref{thm1-2d} and Proposition~\ref{proNonRes}.  The proofs of Propositions~\ref{proInfinite}, \ref{proInfinite-2d}, and \ref{proinstability} are presented in Sections~\ref{SectionInfinite}, \ref{SectionInfinite-2d}, and \ref{SectionInstability}.

%We recall here that $F_*A$ and $F_*\Sigma$ are defined in \eqref{defASigmac}.

%The proof of Proposition~\ref{fundPro}, which is simply based on the change of variables, is left to the reader (see e.g. \cite{KohnOnofreiVogeliusWeinstein}, \cite{NguyenHelmholtz}).

\section{Proof of Theorem~\ref{thm1-3d}}\label{Section-3d}

It is clear from the definition of $(A_c, \Sigma_c)$ in \eqref{defASigmac} that $A_c = {F_{\eps}}_{*}A_\eps$ and $\Sigma_c = {F_{\eps}}_{*}\Sigma_\eps$ where
\begin{equation*}
A_\eps, \Sigma_\eps =
\left\{\begin{array}{cl}
I, 1  & \mbox{if } x \in \mR^3 \setminus B_{\eps}, \\[12pt]
\dsp \frac{1}{\eps} a(x/ \eps), \frac{1}{\eps^3} \sigma(x/ \eps) & \mbox{if } x \in B_{\eps}.
\end{array}\right.
\end{equation*}

Applying Proposition~\ref{fundPro}, Theorem~\ref{thm1-3d} is a consequence of the following
\begin{theorem}\label{thm1-3d*} Let $k > 0$ and  $0< \eps < 1$. Let $u_\eps \in H^1_{\loc}(\mR^3)$ be the unique solution of
\begin{equation*}
\left\{\begin{array}{ll}
\dive (A_\eps \nabla u_\eps) + k^2 \Sigma_\eps u_\eps = {F_\eps^{-1}}_{*} f  & \mbox{ in } \mR^3,\\[12pt]
u_\eps \mbox{ satisfies the outgoing condition}. &
\end{array}\right.
\end{equation*}
In the resonant case, assume in addition that $\dsp \int_{B_1} f \bar \be =0$ for all $\be \in M$. Then for all $r > 0$, there exists a constant $C = C(r, k, a, \sigma)$ which is independent of $\eps$ and $f$ such that
\begin{equation}\label{est-3d}
\|u_\eps - u \|_{H^1(B_{3r} \setminus B_{2r})} \le C \eps \| f \|_{L^2}.
\end{equation}
Moreover, if $U_\eps(x) = u_\eps(\eps x)$, then $U_\eps \rightharpoonup Cl_3(f)$ weakly in $H^1(B_1)$. Consequently,
\begin{equation*}
\lim_{\eps \to 0} \|U_\eps -  Cl_3(f)\|_{L^2(B_1)} = 0.
\end{equation*}
\end{theorem}

Here the operator $Cl_3$ is given in Definition~\ref{defCl3}.

\medskip
The rest of this section will be devoted to the proof of Theorem~\ref{thm1-3d*}.

\subsection{Preliminaries}
\medskip
In this section we present some lemmas which will be used in the proof of Theorem~\ref{thm1-3d*}. Instead of dealing with the unit ball $B_1$, we will present results for a smooth open subset $D$ of $\mR^3$.  We will also suppose that $D \subset B_1$ and $\mR^3 \setminus D$ is connected. We first ``recall" the following result \cite[Lemma 2.2]{NguyenHelmholtz} which will be useful in our analysis.

\begin{lemma} \label{lem0-3d} Let $0 < \eps < 1$, and $g_\eps \in H^{\frac{1}{2}}(\partial D)$. Assume that $v_\eps \in H^1_{\loc}(\mR^3 \setminus \bar D)$ is the unique solution of
\begin{equation*}
\left\{\begin{array}{ll}
\Delta v_\eps + \eps^2  v_\eps = 0 & \mbox{in } \mR^3 \setminus \bar D, \\[12pt]
v_\eps = g_\eps & \mbox{on } \partial D, \\[12pt]
v_\eps \mbox{ satisfies the outgoing condition}.
\end{array}\right.
\end{equation*}

\begin{enumerate}
\item[$i)$] We have
\begin{equation}\label{state1-3d}
\| v_\eps \|_{H^1(B_r \setminus \overline D)} \le C_r \| g_\eps\|_{H^\frac{1}{2}(\partial D)} \quad \forall \, r > 5,
\end{equation}
and
\begin{equation}\label{state2-3d}
\eps^{1/2} \| v_\eps \|_{L^2(B_{4r/\eps} \setminus B_{r/ \eps})} \le C_r \| g_\eps \|_{H^\frac{1}{2}(\partial D)},
\end{equation}
for some positive constants $C_r =C(r, D)$.

\item[$ii)$] Assume in addition that $g_\eps \rightharpoonup g $ weakly in $H^\frac{1}{2}(\partial D)$ as $\eps \to 0$. Then $v_\eps \rightharpoonup v$ weakly in $H^1_{\loc}(\mR^3 \setminus \overline D)$ where $v \in W^1(\mR^3 \setminus \overline D)$ \footnote{The space $W^1$ is defined in \eqref{defW13d}.} is the unique solution of
\begin{equation}\label{sysv}
\left\{\begin{array}{ll}
\Delta v = 0 & \mbox{ in } \mR^3 \setminus \overline D, \\[12pt]
v = g & \mbox{ on } \partial D.
\end{array}\right.
\end{equation}
\end{enumerate}

\end{lemma}

\noindent {\bf Proof.} Inequalities \eqref{state1-3d}, and \eqref{state2-3d} with $r =1$ are in \cite[Lemma 2.2]{NguyenHelmholtz} \footnote{There is a typo in \cite[(2.4)]{NguyenHelmholtz} for $d=3$ where the term $\eps^{1/2}$ is missing} . They are consequences of the fact that the fundamental solution of the Helmholtz equation converges to the fundamental solution of the Laplace equation as the frequency goes to 0 in $3d$. The proof of \eqref{state2-3d} in the general case follows in the same manner. Part $ii)$ follows from part $i)$ as follows. From part $i)$, one may assume that $v_\eps \rightharpoonup v$ weakly in $H^1_{\loc}(\mR^3 \setminus \overline D)$ (up to a subsequence) and $v$ satisfies \eqref{sysv}. Using the representation formula and the the fact that the fundamental solution of the Helmholtz equation converges to the fundamental solution of the Laplace equation as the frequency goes to 0 in $3d$, one can prove that $v \in W^1(\mR^3 \setminus \overline D)$. Since \eqref{sysv} has a unique solution $v \in W^1(\mR^3 \setminus \overline D)$ (see e.g. \cite[Theorem 2.5.14 on page 64]{Nedelec} ), the result holds for the whole family $(v_\eps)$. The details of the proof are left to the reader. \proofend

\medskip

In what follows,  $a$ denotes a real symmetric matrix-valued function and $\sigma$ denotes a complex function defined on $D$. We also assume that $a$ is uniformly elliptic and $\sigma$ satisfies $0< \ess \inf \Re \sigma \le \ess \sup \Re \sigma < + \infty$ and $0 \le \ess \inf \Im \sigma \le \ess \sup \Im \sigma < + \infty$.
We define
\begin{equation}\label{defM}
M := \Big\{\psi \in H^1(D); \; \dive (a \nabla \psi) + k^2 \sigma \psi = 0 \mbox{ in } D \mbox{ and } a \nabla \psi \cdot \eta = 0 \mbox{ on } \partial D \Big\}.
\end{equation}

\medskip
The following lemma establishes the uniqueness of $(v, w)$ in Definition~\ref{defCl3}. This lemma is also used in the proof of Lemma~\ref{lemd=3}.

\begin{lemma}\label{lemUniqueness-3d} Assume that the system is resonant i.e.  $M \neq \{ 0\}$.  Then there exists no nonzero solution $(v,w)$ in $W^1(\mR^3)  \times M^{\perp}$ of the systems
\begin{equation}\label{sysvw}\left\{
\begin{array}{ll}
\Delta v  = 0 & \mbox{ in } \mR^3 \setminus \overline D, \\[12pt]
\dive(a \nabla  v) + k^2 \sigma v = 0 & \mbox{ in } D, \\[12pt]
\dive(a \nabla  w) + k^2 \sigma w = 0 & \mbox{ in } D,
\end{array}\right. \mbox{  and  }
\left\{
\begin{array}{ll}
\dsp a \nabla v \cdot \eta \Big|_{\mathrm{int}} = 0 & \mbox{ on } \partial D, \\[12pt]
\dsp \frac{\partial v}{\partial \eta} \Big|_{\mathrm{ext}} - a \nabla w \cdot \eta = 0 & \mbox{ on } \partial D.
\end{array}\right.
\end{equation}
Here $M$ and $M^{\perp}$ are respectively defined in \eqref{defM}  and  \eqref{defMorth}.
\end{lemma}

\noindent{\bf Proof.} Since $v \in W^1(\mR^3 \setminus \overline D)$, it follows from \eqref{sysvw} that
\begin{equation}\label{nablav1}
\int_{\mR^3 \setminus D} |\nabla v|^2  = - \int_{\partial D} \frac{\partial v}{\partial \eta} \Big|_{\mathrm{ext}}\bar v = - \int_{\partial D} (a \nabla w \cdot \eta ) \bar v.
\end{equation}
On the other hand, from \eqref{sysvw}, we have
\begin{equation*}
\int_{D} \langle a\nabla v, \nabla v \rangle - \int_{D} k^2 \sigma |v|^2 =0.
\end{equation*}
This implies that $v =0$ on the set $\{ \Im \sigma > 0 \}$. Thus we deduce from  \eqref{sysvw} that
\begin{equation}\label{nablav3-1}
\int_{\partial D} (a \nabla w \cdot \eta ) \bar v = \int_{D} a \nabla w \nabla  \bar v - \int_{D} k^2 \sigma w \bar v \\[12pt]
= \overline{\int_{D} a \nabla v \nabla  \bar w - \int_{D} k^2 \sigma v \bar w}
\end{equation}
and
\begin{equation}\label{nablav3-2}
\int_{D} a \nabla v \nabla  \bar w - \int_{D} k^2 \sigma v \bar w  = \int_{\partial D} (a \nabla v \cdot \eta \Big|_{\mathrm{int}}) \bar w = 0.
\end{equation}
A combination of \eqref{nablav1}, \eqref{nablav3-1}, and \eqref{nablav3-2} yields $v = 0$ in $\mR^3 \setminus D$. It follows that, by \eqref{sysvw} and the unique continuation principle,
\begin{equation*}
v = 0 \quad \mbox{ in } \mR^3
\end{equation*}
From \eqref{sysvw},  $a \nabla w \cdot \eta = 0$. Hence $w = 0$ since $w \in M^{\perp}$ and $\dive(a \nabla w) + k^2 \sigma w =0$. \proofend

\medskip We now establish the crucial ingredient in the proof of Theorem~\ref{thm1-3d*}.

\begin{lemma}\label{lemd=3} Let $0 < \eps < 1$ and $k> 0$. Let $\theta_\eps \in L^2(D)$, $g_\eps \in H^{-\frac{1}{2}}(\partial D)$, and $v_\eps \in H^1_{\loc}(\mR^3)$ be the unique solution of the system
\begin{equation*}
\left\{\begin{array}{ll}
\Delta v_\eps + \eps^2 k^2 v_\eps = 0 & \mbox{in } \mR^3 \setminus \bar D, \\[12pt]
\dive  (a \nabla v_\eps) + k^2 \sigma v_\eps = \theta_\eps & \mbox{in } D, \\[12pt]
\dsp \frac{\partial v_\eps}{\partial \eta}\Big|_{\mathrm{ext}} - \frac{1}{\eps} a \nabla v_\eps \cdot \eta \Big|_{\mathrm{int}} = g_\eps & \mbox{on } \partial D \\[12pt]
v_\eps \mbox{ satisfies the outgoing condition}.
\end{array}\right.
\end{equation*}
In the case $M \neq \{0\}$ \footnote{$M$ is defined in \eqref{defM}.}, assume in addition that $\dsp \int_{B_1} \theta_\eps \bar \be = 0 $ for all $\be \in M$. We have
\begin{equation}\label{estveps}
\|v_\eps \|_{H^1(B_5)} \le C \big( \|\theta_\eps \|_{L^2(D)} +  \| g_\eps \|_{H^{-\frac{1}{2}}(\partial D)} \big),
\end{equation}
for some positive constant $C$ depending only on $k$, $a$, $\sigma$, and $D$ but independent of $\eps$, $\theta_\eps$ and $g_\eps$.
Assume that $\theta_\eps \rightharpoonup \theta $ weakly in $L^2(D)$ and  $g_\eps \rightharpoonup g$ weakly in $H^{-\frac{1}{2}}(\partial D)$ as $\eps \to 0$. Then $v_\eps \rightharpoonup v$ weakly in $H^1(D)$. Consequently,
\begin{equation*}
\lim_{\eps \to 0}\|v_\eps - v\|_{L^2(D)} = 0.
\end{equation*}
Here
\begin{enumerate}
\item[$i)$] in the non-resonant case: $M = \{ 0\}$, $v \in H^1(D)$ is the unique solution of the system
\begin{equation}\label{sysvNon-R}\left\{
\begin{array}{ll}
\dive(a \nabla  v) + k^2 \sigma v = \theta & \mbox{ in } D, \\[12pt]
\dsp a \nabla v \cdot \eta  = 0 & \mbox{ on } \partial D.
\end{array}\right.
\end{equation}

\item[$ii)$] in the resonant case: $M \neq \{ 0\}$, $v$ is the first component of the pair $(v, w) \in W^1(\mR^3) \times M^{\perp}$ which is the unique solution of the systems
\begin{equation}\label{sysvR}\left\{
\begin{array}{ll}
\Delta v  = 0 & \mbox{ in } \mR^3 \setminus \overline D, \\[12pt]
\dive(a \nabla  v) + k^2 \sigma v = \theta & \mbox{ in } D, \\[12pt]
\dive(a \nabla  w) + k^2 \sigma w = 0 & \mbox{ in } D,
\end{array}\right. \quad \mbox{and} \quad
\left\{ \begin{array}{ll}
\dsp a \nabla v \cdot \eta \Big|_{\mathrm{int}} = 0 & \mbox{ on } \partial D, \\[12pt]
\dsp \frac{\partial v}{\partial \eta} \Big|_{\mathrm{ext}} - a \nabla w \cdot \eta = g & \mbox{ on } \partial D.
\end{array}\right.
\end{equation}
Here  $M^{\perp}$ is defined by  \eqref{defMorth}.
\end{enumerate}
\end{lemma}

\begin{remark} The uniqueness of $(v, w)$ follows from Lemma~\ref{lemUniqueness-3d}. The existence of $(v, w)$ will be proved in the proof of Lemma~\ref{lemd=3}.
\end{remark}

\noindent{\bf Proof.} We first prove that
\begin{equation}\label{conclu1d=3}
\|v_\eps \|_{L^2(B_5)} \le C \big( \| \theta_\eps\|_{L^2(D)} + \| g_\eps \|_{H^{-\frac{1}{2}}(\partial D)} \big),
\end{equation}
by contradiction. Suppose that  there exist $(\theta_n) \subset L^2(D)$ (we also assume that $\dsp \int_{D}\theta_n \bar \be = 0 $ for any $ \be \in M$ in the resonant case),  $(g_n) \subset H^{-\frac{1}{2}}(\partial D)$ and $\eps_n \to 0$ such that
\begin{equation}\label{contradiction1}
\|v_n \|_{L^2(B_5)}  = 1 \quad \mbox{and} \quad \lim_{n \lr \infty} \big( \| \theta_n\|_{L^2(D)} + \| g_n \|_{H^{-\frac{1}{2}}(\partial D)}\big) = 0.
\end{equation}
Here $v_n \in H^1_{\loc}(\mR^3)$ is the unique solution of
\begin{equation}\label{sysvnd=3}
\left\{\begin{array}{ll}
\Delta v_n + \eps_n^2 k^2 v_n = 0 & \mbox{in } \mR^3 \setminus \bar D, \\[12pt]
\dive (a \nabla v_n) + k^2 \sigma v_n = \theta_n & \mbox{in } D, \\[12pt]
\dsp \frac{\partial v_n}{\partial \eta}\Big|_{\mathrm{ext}} - \frac{1}{\eps_n}  a \nabla v_n \cdot \eta  \Big|_{\mathrm{int}} = g_n & \mbox{on } \partial D \\[12pt]
v_n \mbox{ satisfies the outgoing condition}.
\end{array}\right.
\end{equation}
Multiplying system~\eqref{sysvnd=3} by $\bar v_n$ (the conjugate of $v_n$) and integrating the expression obtained over $B_4$, we have
\begin{multline}\label{sysvnd=3-1.1}
\int_{B_4 \setminus D} |\nabla v_n|^2 - \eps_n^2 k^2 \int_{B_4 \setminus D} |v_n|^2 + \frac{1}{\eps_n} \int_{D} \langle a \nabla v_n, \nabla v_n \rangle - \frac{1}{\eps_n} \int_{D} k^2 \sigma |v_n|^2  \\
=   \int_{\partial B_4} \frac{\partial v_n}{\partial r} \bar v_n - \int_{\partial D} g_n \bar v_n - \frac{1}{\eps_n} \int_{D} \theta_n \bar v_n.
\end{multline}
Applying Lemma~\ref{lem0-3d}, we deduce from \eqref{contradiction1} and \eqref{sysvnd=3-1.1} that
\begin{equation*}
\int_{D} |\nabla v_n|^2  \le C,
\end{equation*}
which implies, since $\|v_n \|_{L^2(D)} \le 1$,
\begin{equation} \label{estGradvnInt}
\| v_n \|_{H^\frac{1}{2}(\partial D)} \le C.
\end{equation}
Using Lemma~\ref{lem0-3d}, from \eqref{contradiction1}, \eqref{sysvnd=3}, and \eqref{estGradvnInt}, we have
\begin{equation}\label{estGradvn}
\|v_n \|_{H^1(B_r)} \le C_r \quad \forall \, r > 0.
\end{equation}

\noindent \underline{Case 1:} The non-resonant case: $M = \{ 0\}$.

\medskip
From \eqref{estGradvn} and part $ii)$ of Lemma~\ref{lem0-3d}, one may assume that $v_n \rightharpoonup v$ weakly in $H^1_{\loc}(\mR^3)$, $v_n \to v$ in $L^2(B_5)$ for some $v  \in W^{1}(\mR^3)$ such that
\begin{equation*}\left\{
\begin{array}{ll}
\Delta v  = 0 & \mbox{ in } \mR^3 \setminus \overline D, \\[12pt]
\dive (a \nabla v)  + k^2 \sigma v = 0 & \mbox{ in } D, \\[12pt]
a \nabla v \cdot \eta \Big|_{\mathrm{int}} = 0 & \mbox{ on } \partial D.
\end{array}\right.
\end{equation*}
Since $M = \{ 0\}$, it follows that
\begin{equation}\label{v=0-1}
v=0 \quad \mbox { in } D.
\end{equation}
Hence
\begin{equation*}\left\{
\begin{array}{ll}
\Delta v  = 0 & \mbox{ in } \mR^3 \setminus \overline D, \\[12pt]
v = 0 & \mbox{ on } \partial D.
\end{array}\right.
\end{equation*}
Since $v \in W^1(\mR^3 \setminus \overline D)$ (by Lemma~\ref{lem0-3d}), it follows that (see e.g. \cite[Theorem 2.5.14 on page 64]{Nedelec})
\begin{equation}\label{v=0-2}
v = 0 \quad \mbox{ in } \mR^3 \setminus D.
\end{equation}
Combining \eqref{v=0-1} and \eqref{v=0-2} yields $v =0$. We have a contradiction since $\|v \|_{L^2(B_5)} = \lim_{n \to \infty} \| v_n\|_{L^2(B_5)}=1$.
\medskip

\noindent \underline{Case 2:} The resonant case: $M \not = \{ 0 \}$.

\medskip
Since $\dsp \int_{D} \theta_n \bar \be = 0$ for any $\be \in M$, it follows from  the Fredholm alternative theory (see e.g. \cite{BrAnalyse} or \cite{LaxFunct}) that  there exists a unique solution $v_{1, n} \in M^{\perp}$  of the system
\begin{equation}\label{sysv1n*}\left\{
\begin{array}{ll}
\dive (a \nabla v_{1,n})  + k^2 \sigma v_{1, n} = \theta_n & \mbox{ in } D, \\[12pt]
a \nabla v_{1, n} \cdot \eta  = 0 & \mbox{ on } \partial D.
\end{array}\right.
\end{equation}
Moreover, we have
\begin{equation}\label{estv1n}
\| v_{1, n}\|_{H^1(D)} \le C \| \theta_n\|_{L^2(D)},
\end{equation}
for some positive constant $C$ independent of $\theta_n$. Let $v_{2,n}$ be the projection of $v_n - v_{1, n}$ into $M$ i.e. $v_{2, n} \in M$ and $v_{n} - v_{1, n} - v_{2, n} \in M^{\perp}$. Set
\begin{equation}\label{defwn-1}
w_{n} = \frac{1}{\eps_n}(v_n - v_{1, n} - v_{2, n}) \;  (\in M^{\perp}).
\end{equation}
Then
\begin{equation}\label{defwn}
a \nabla v_n \cdot \eta \Big|_{\mathrm{int}} = \eps_n a \nabla w_n \cdot \eta \quad \mbox{ on } \partial D.
\end{equation}
We deduce from \eqref{sysvnd=3} and the definition of $M$ that
\begin{equation}\label{v2n-partialD}\left\{
\begin{array}{cl}
\dive (a \nabla w_n) + k^2 \sigma w_n = 0 & \mbox{ in } D, \\[12pt]
\dsp a \nabla w_{n} \cdot \eta  = \frac{\partial v_{n}}{\partial \eta} \Big|_{\mathrm{ext}} - g_n & \mbox{ on } \partial D.
\end{array}\right.
\end{equation}
Combining \eqref{contradiction1}, \eqref{estGradvn}, and \eqref{v2n-partialD} yields
\begin{equation*}
\| a \nabla w_{n} \cdot \eta  \|_{H^{-\frac{1}{2}}(\partial D)} \le \Big\| \frac{\partial v_{n}}{\partial \eta} \Big|_{\mathrm{ext}} \Big\|_{H^{-\frac{1}{2}}(\partial D)} + \| g_n\|_{H^{-\frac{1}{2}}(\partial D)} \le C.
\end{equation*}
Since $\dive (a \nabla w_{n}) + k^2 \sigma w_{n} = 0 $ in $D$ and $w_{n} \in M^{\perp}$, it follows that
\begin{equation}\label{prown}
\|w_{n}\|_{H^1(D)} \le C.
\end{equation}
From \eqref{estGradvn}, \eqref{prown}, and the fact that $w_n  \in M^{\perp}$, one may assume that
\begin{equation*}
\left\{\begin{array}{l}
v_{n} \rightharpoonup v \mbox{ weakly in } H^1_{\loc}(\mR^3),\\[12pt]
w_{n} \rightharpoonup w  \in M^{\perp} \mbox{ weakly in } H^1 (D).
\end{array} \right.
\end{equation*}
By Lemma~\ref{lem0-3d}, it follows from \eqref{contradiction1}, \eqref{sysvnd=3}, \eqref{defwn}, and \eqref{v2n-partialD} that $(v, w) \in W^1(\mR^3) \times M^{\perp}$ is a solution of the system
\begin{equation*}\left\{
\begin{array}{ll}
\Delta v  = 0 & \mbox{ in } \mR^3 \setminus \overline D, \\[12pt]
\dive(a \nabla  v) + k^2 \sigma v = 0 & \mbox{ in } D, \\[12pt]
\dive(a \nabla  w) + k^2 \sigma w = 0 & \mbox{ in } D,
\end{array}\right. \quad \mbox{and} \quad
\left\{
\begin{array}{ll}
\dsp a \nabla v \cdot \eta \Big|_{\mathrm{int}} = 0 & \mbox{ on } \partial D, \\[12pt]
\dsp \frac{\partial v}{\partial \eta} \Big|_{\mathrm{ext}} = a \nabla w \cdot \eta & \mbox{ on } \partial D.
\end{array}\right.
\end{equation*}
We deduced from Lemma~\ref{lemUniqueness-3d} that $v=0$. We have a contradiction since $\| v\|_{L^2(B_5)} =1$.

\medskip

From Cases 1 and 2, \eqref{conclu1d=3} is proved. Hence we obtain \eqref{estveps} by using \eqref{conclu1d=3} and applying the same arguments used to get \eqref{estGradvn}. The conclusion of second part is a consequence of \eqref{estveps} and can be processed as follows. From \eqref{estveps}, one may assume that $v_\eps \rightharpoonup v$ weakly in $H^1(D)$ (up to a subsequence). In the non-resonant case, $v \in H^1(D)$ is a solution of the system \eqref{sysvNon-R}. Since \eqref{sysvNon-R} has a unique solution $v \in H^1(D)$, the conclusion in this case holds for the whole family. In the resonant case, applying the same decomposition as in~\eqref{defwn-1} and using the same facts as in \eqref{sysvnd=3}, \eqref{defwn}, and \eqref{v2n-partialD}, one may assume that (up to a subsequence)
\begin{equation*}
\left\{\begin{array}{l}
v_{\eps} \rightharpoonup v  \in W^1(\mR^3) \mbox{ weakly in } H^1_{\loc}(\mR^3),\\[12pt]
w_{\eps} \rightharpoonup w  \in M^{\perp} \mbox{ weakly in } H^1 (D),
\end{array} \right.
\end{equation*}
and $(v, w) \in W^1(\mR^3) \times M^{\perp}$ satisfies system \eqref{sysvR}. Since system \eqref{sysvR} has a unique solution in $(v, w) \in W^1(\mR^3) \times M^{\perp}$ (by Lemma~\ref{lemUniqueness-3d}), the conclusion holds for the whole family.  \proofend

\subsection{\bf Proof of Theorem~\ref{thm1-3d*}.}

We follow the method used in the proof of \cite[Theorem 2.1]{NguyenHelmholtz}. Let $u_{1, \eps} \in H^1_{\loc}(\mR^3)$ be the unique solution of
\begin{equation*}
\left\{\begin{array}{ll}
\Delta u_{1, \eps} + k^2 u_{1, \eps} = f ~(= {F_\eps^{-1}}_{*} f) & \mbox{ in } \mR^3 \setminus \overline B_\eps,\\[12pt]
u_{1, \eps} = 0 & \mbox{ in } B_\eps \\[12pt]
u_{1, \eps} \mbox{ satisfies the outgoing condition}. &
\end{array}\right.
\end{equation*}
Define
\begin{equation*}
w_{1, \eps} = u_{1, \eps} - u,
\end{equation*}
and
\begin{equation*}
w_{2, \eps} = u_\eps - u_{1, \eps}.
\end{equation*}
We claim that
\begin{equation}\label{claim-w1}
\|w_{1, \eps} \|_{H^1(B_{4r} \setminus B_{r})} \le C_r \eps \| f \|_{L^2}
\end{equation}
and
\begin{equation}\label{claim-w2}
 \|w_{2, \eps} \|_{H^1(B_{4r} \setminus B_{r})} \le C_r \eps \| f \|_{L^2}.
\end{equation}

\medskip
\noindent\underline{Proof of Claim~\eqref{claim-w1}}. From the definition of $w_{1, \eps}$, it follows that $w_{1, \eps} \in H^1_{\loc}(\mR^3)$ and $w_{1, \eps}$ satisfies
\begin{equation}
\left\{\begin{array}{ll}
\Delta w_{1, \eps} + k^2 w_{1, \eps} = 0 & \mbox{ in } \mR^3 \setminus \overline B_\eps,\\[12pt]
w_{1, \eps} = - u & \mbox{ in } \partial B_\eps \\[12pt]
w_{1, \eps} \mbox{ satisfies the outgoing condition}. &
\end{array}\right.
\end{equation}
Define $W_{1, \eps}(x) = w_{1, \eps}(\eps x)$. Then $W_{1, \eps} \in H^1_{\loc}(\mR^3)$ and $W_{1, \eps}$ satisfies
\begin{equation}\label{defW1eps}
\left\{\begin{array}{ll}
\Delta W_{1, \eps} + \eps^2 k^2 W_{1, \eps} = 0 & \mbox{ in } \mR^3 \setminus \overline B_1,\\[12pt]
W_{1, \eps} = -  u (\eps \cdot) & \mbox{ in } \partial B_1 \\[12pt]
W_{1, \eps} \mbox{ satisfies the outgoing condition}. &
\end{array}\right.
\end{equation}
Since $\| u (\eps \cdot)\|_{H^\frac{1}{2} (\partial B_1)} \le \| f\|_{L^2}$, by Lemma~\ref{lem0-3d}, we have
\begin{equation*}
\eps \int_{B_{4r/ \eps} \setminus B_{r/ \eps}} |W_{1, \eps}|^2 \le C_r \| f\|_{L^2}^2.
\end{equation*}
By a change of variables,
\begin{equation}\label{reg1}
\int_{B_{4r} \setminus B_{r}} |w_{1, \eps}|^2 \le C_r \eps^2 \| f\|_{L^2}^2 .
\end{equation}
Since $\Delta w_{1, \eps} + k^2 w_{1, \eps} = 0$ in $\mR^3 \setminus \bar B_{\eps}$, by the regularity theory of elliptic equations,  Claim~\eqref{claim-w1} follows from \eqref{reg1}.

\medskip
\noindent\underline{Proof of Claim~\eqref{claim-w2}}. It is clear that $w_{2, \eps} \in H^1_{\loc}(\mR^3)$ is the unique solution of
\begin{equation*}
\left\{\begin{array}{ll}
\Delta w_{2, \eps}  +  k^2 w_{2, \eps} = 0 & \mbox{ in } \mR^3 \setminus \overline B_\eps, \\[12pt]
\dive  (A_\eps \nabla w_{2, \eps}) + k^2 \Sigma_\eps w_{2, \eps} = {F_\eps^{-1}}_{*} f & \mbox{ in } B_\eps, \\[12pt]
\dsp \frac{\partial w_{2, \eps}}{\partial \eta} \Big|_{\mathrm{ext}} - A_\eps \nabla w_{2, \eps} \cdot \eta \Big|_{\mathrm{int}}= - \frac{\partial u_{1, \eps}}{\partial \eta} & \mbox{ on } \partial B_\eps, \\[12pt]
w_{2, \eps} \mbox{ satisfies the outgoing condition.} &
\end{array}\right.
\end{equation*}
Define $W_{2, \eps}(x) = w_{2, \eps}(\eps x)$. Then $W_{2, \eps} \in H^1_{\loc}(\mR^3)$ and $W_{2, \eps}$ is the unique solution of
\begin{equation*}
\left\{\begin{array}{ll}
\Delta W_{2, \eps} + \eps^2 k^2 W_{2, \eps} = 0 & \mbox{in } \mR^3 \setminus \overline B_1, \\[12pt]
\dive (a \nabla W_{2, \eps}) + k^2 \sigma W_{2, \eps} = f & \mbox{in } B_1, \\[12pt]
\dsp \frac{\partial W_{2, \eps}}{\partial \eta}\Big|_{\mathrm{ext}} - \frac{1}{\eps} a \nabla  W_{2, \eps} \cdot \eta \Big|_{\mathrm{int}} = - \eps \frac{\partial u_{1, \eps}}{\partial \eta} (\eps x) & \mbox{on } \partial B_1, \\[12pt]
W_{2, \eps} \mbox{ satisfies the outgoing condition}. &
\end{array}\right.
\end{equation*}
Since  $u_{1, \eps} = w_{1, \eps} + u$ and $W_{1, \eps} = w_{1, \eps} (\eps \cdot)$, it follows that
\begin{equation}\label{tt1}
\eps \frac{\partial u_{1, \eps}}{\partial \eta} (\eps x) = \frac{\partial W_{1, \eps}}{\partial \eta}(x) + \eps \frac{\partial u}{\partial \eta} (\eps x) \quad \mbox{ on } \partial B_1.
\end{equation}
From \eqref{defW1eps} and Lemma~\ref{lem0-3d}, we obtain
\begin{equation*}
\Big\|\eps \frac{\partial u_{1, \eps}}{\partial \eta} (\eps x) \Big\|_{H^{-\frac{1}{2}}(\partial B_1)} \le C \| f\|_{L^2}.
\end{equation*}
We deduce from Lemma~\ref{lemd=3} that
\begin{equation*}
\| W_{2, \eps}\|_{H^1(B_5)} \le C \| f\|_{L^2}.
\end{equation*}
Applying Lemma~\ref{lem0-3d}, we have
\begin{equation*}
\eps \int_{B_{4r/\eps} \setminus B_{r/\eps}} |W_{2, \eps}|^2 \le C_r \| f\|_{L^2}^2.
\end{equation*}
By a change of variables,
\begin{equation*}
\int_{B_{4r} \setminus B_{r}} |w_{2, \eps}|^2 \le C_r \eps^2 \| f\|_{L^2}^2.
\end{equation*}
Claim~\eqref{claim-w2} now follows from the regularity theory of elliptic equations and the fact that $\Delta w_{2, \eps} + k^2 w_{2, \eps} = 0$ in $\mR^3 \setminus \bar B_\eps$.

\medskip
Thus Claims~\eqref{claim-w1} and \eqref{claim-w2} are proved.  Since $u_\eps - u = w_{1, \eps} + w_{2, \eps}$, \eqref{est-3d} follows.

\medskip
The rest of the proof goes as follows. Since $u(\eps \cdot) \lr u(0)$, it follows from \eqref{defW1eps} and Lemma~\ref{lem0-3d} that  $W_{1, \eps} \rightharpoonup W_1$ weakly in $H^1_{\loc}(\mR^3 \setminus \overline B_1)$ where $W_1 \in W^1(\mR^3 \setminus \overline B_1)$ is the unique solution of
\begin{equation}\label{sysW1}
\left\{\begin{array}{ll}
\Delta W_1 = 0 & \mbox{ in } \mR^3 \setminus \overline B_1,\\[12pt]
W_1 = -  u (0) & \mbox{ in } \partial B_1.
\end{array}\right.
\end{equation}

\medskip
\noindent \underline{Case 1}: The non-resonant case: $M = \{ 0 \}$. By Lemma~\ref{lemd=3}, $W_{2, \eps} \rightharpoonup \hat v$ weakly in $H^1(B)$ where $\hat v$ is the unique solution of the system
\begin{equation*}\left\{
\begin{array}{ll}
\dive(a \nabla  \hat v) + k^2 \sigma \hat v = f & \mbox{ in } B_1, \\[12pt]
\dsp a \nabla \hat v \cdot \eta \Big|_{\mathrm{int}} = 0 & \mbox{ on } \partial B_1.
\end{array}\right.
\end{equation*}
Therefore, the conclusion follows in this case.

\medskip
\noindent \underline{Case 2:} The resonant case: $M \neq \{ 0 \}$. By Lemma~\ref{lemd=3}, it follows from \eqref{tt1} that $W_{2, \eps} \rightharpoonup \hat v$ weakly in $H^1_{\loc}(\mR^3)$ where $\hat v$ is the first component of the pair $(\hat v, \hat w) \in W^1(\mR^3) \times M^\perp$ which is the unique solution of the systems
    \begin{equation}\label{syshat}\left\{
\begin{array}{ll}
\Delta \hat v  = 0 & \mbox{ in } \mR^3 \setminus \overline B_1, \\[12pt]
\dive(a \nabla  \hat v) + k^2 \sigma \hat v = f & \mbox{ in } B_1, \\[12pt]
\dive(a \nabla  \hat w) + k^2 \sigma \hat w = 0 & \mbox{ in } B_1,
\end{array}\right. \mbox{ and }  \left\{
\begin{array}{ll}
\dsp a \nabla \hat v \cdot \eta \Big|_{\mathrm{int}} = 0 & \mbox{ on } \partial B_1, \\[12pt]
\dsp \frac{\partial \hat v}{\partial \eta} \Big|_{\mathrm{ext}} + \frac{\partial W_1}{\partial \eta}= a \nabla \hat w \cdot \eta & \mbox{ on } \partial B_1.
\end{array}\right.
\end{equation}
Hence $\hat v$ depends on $u(0)$ through $W_1$ (see the transmission condition in \eqref{syshat} and the system of $W_1$ in \eqref{sysW1}). Define $v_{\mathrm{ext}} = \hat v + W_1$ for $x \in \mR^3 \setminus \overline B_1$, $v_{\mathrm{int}} = \hat v$ if $x \in B_1$, and $w = \hat w$ if $x \in B_1$. It follows from \eqref{sysW1} and \eqref{syshat} that the triple $(v_{\mathrm{ext}}, v_{\mathrm{int}}, w) \in W^1(\mR^3 \setminus \overline B_1) \times H^1(B_1) \times M^\perp$ is the unique solution of the systems
    \begin{equation*}\left\{
\begin{array}{ll}
\Delta v_{\mathrm{ext}}  = 0 & \mbox{ in } \mR^3 \setminus \overline B_1, \\[12pt]
\dive(a \nabla  v_{\mathrm{int}}) + k^2 \sigma v_{\mathrm{int}} = f & \mbox{ in } B_1, \\[12pt]
\dive(a \nabla  w) + k^2 \sigma w = 0 & \mbox{ in } B_1,
\end{array}\right. \mbox{ and }  \left\{
\begin{array}{ll}
\dsp v_{\mathrm{ext}} - v_{\mathrm{int}} = - u(0) & \mbox{ on } \partial B_1, \\[12pt]
\dsp a \nabla v_{\mathrm{int}} \cdot \eta = 0 & \mbox{ on } \partial B_1, \\[12pt]
\dsp \frac{\partial v_{\mathrm{ext}}}{\partial \eta}  = a \nabla w \cdot \eta & \mbox{ on } \partial B_1.
\end{array}\right.
\end{equation*}
Since $U_\eps = W_{2, \eps}$ in $B_1$ and $W_{2, \eps} \rightharpoonup \hat v = v_{\mathrm{int}} = Cl_3(f)$ in $H^1(B_1)$, we obtain the conclusion in this case.
\proofend

\section{Proofs of Theorem~\ref{thm1-2d} and Proposition~\ref{proNonRes}} \label{Section-2d}

According to Proposition~\ref{fundPro}, Theorem~\ref{thm1-2d} is a consequence of the following

\begin{theorem}\label{thm1-2d*} Let $d=2$, $k > 0$, and $0< \eps < 1$. Define
\begin{equation*}
A_\eps, \Sigma_\eps =
\left\{\begin{array}{cl}
I, 1  & \mbox{if } x \in \mR^2 \setminus B_{\eps}, \\[12pt]
\dsp a(x/ \eps), \frac{1}{\eps^2} \sigma(x/ \eps) & \mbox{if } x \in B_{\eps}.
\end{array}\right.
\end{equation*}
Let $u_\eps \in H^1_{\loc}(\mR^2)$ be the unique solution of
\begin{equation*}
\left\{\begin{array}{ll}
\dive (A_\eps \nabla u_\eps) + k^2 \Sigma_\eps u_\eps = F_\eps^{-1} * f & \mbox{ in } \mR^2,\\[12pt]
u_\eps \mbox{ satisfies the outgoing condition}. &
\end{array}\right.
\end{equation*}
Assume the system is non-resonant \footnote{The non-resonant system is defined in Definition~\ref{defCl2}}. We have
\begin{equation}\label{est-2d}
\|u_\eps - u \|_{H^1(B_{4r} \setminus B_{r})} \le \frac{C}{|\ln \eps|} \| f \|_{L^2},
\end{equation}
for some $C = C(r, k, a, \sigma)$.  Moreover, if  $U_\eps(x) = u_\eps(\eps x)$, then $U_\eps \rightharpoonup Cl_2(f)$ weakly in $H^1(B_1)$. Consequently,
\begin{equation*}
\lim_{\eps \to 0} \|U_\eps - Cl_2(f)\|_{L^2(B_1)} = 0.
\end{equation*}
\end{theorem}

In the rest of this section, we present the proofs of Theorem~\ref{thm1-2d*} and Proposition~\ref{proNonRes}.

\subsection{Preliminaries}

In this section we present some lemmas which will be used in the proof of Theorem~\ref{thm1-2d*}. Instead of dealing with the unit ball $B_1$, we will present results for a smooth open subset $D$ of $\mR^2$.  We will also assume that $D \subset B_1$ and $\mR^2 \setminus D$ is connected. We first ``recall" the following result \cite[Lemma 2.2]{NguyenHelmholtz}  which will be useful in our analysis.

\begin{lemma} \label{lem0-2d} Let $0 < \eps < 1$, and $g_\eps \in H^{\frac{1}{2}}(\partial D)$. Assume that $\mR^2 \setminus D$ is connected and $v_\eps \in H^1_{\loc}(\mR^2)$ is the unique solution of
\begin{equation*}
\left\{\begin{array}{ll}
\Delta v_\eps + \eps^2  v_\eps = 0 & \mbox{in } \mR^2 \setminus \overline D, \\[12pt]
v_\eps = g_\eps & \mbox{on } \partial D \\[12pt]
v_\eps \mbox{ satisfies the outgoing condition}.
\end{array}\right.
\end{equation*}

\begin{enumerate}
\item[$i)$] We have
\begin{equation}\label{state1-2d}
\| v_\eps \|_{H^1(B_r \setminus D)} \le C_r \| g_\eps\|_{H^\frac{1}{2}(\partial D)} \quad \forall \, r > 5,
\end{equation}
and
\begin{equation}\label{state2-2d}
\| v_\eps \|_{L^2(B_{4r/\eps} \setminus B_{r/ \eps})} \le \frac{C_r}{ \eps |\log \eps|} \| g_\eps \|_{H^\frac{1}{2}(\partial D)},
\end{equation}
for some positive constants $C_r =C(r, k, D)$.

\item[$ii)$] Assume that $g_\eps \rightharpoonup g$ weakly in $H^\frac{1}{2}(\partial D)$. Then $v_\eps \rightharpoonup v$ weakly in $H^1_{\loc}(\mR^2)$, where $v \in W^1(\mR^2 \setminus D)$ is the unique solution of
\begin{equation}\label{sysv2d}
\left\{\begin{array}{ll}
\Delta v = 0 & \mbox{ in } \mR^2 \setminus \overline D, \\[12pt]
v = g & \mbox{ on } \partial D.
\end{array}\right.
\end{equation}
\end{enumerate}
\end{lemma}

\noindent{\bf Proof.} Inequalities \eqref{state1-2d}, and \eqref{state2-2d} with $r=1$ are in \cite[Lemma 2.2]{NguyenHelmholtz} \footnote{There is a typo in \cite[(2.4)]{NguyenHelmholtz} for $d=2$ where the term $\eps^{1/2}$ must be replaced by $\eps$.}. The proof of \eqref{state2-2d} in the general case follows in the same manner. To prove part $ii)$ we process as follows. Since $(v_\eps)$ is bounded in $H^1_{\loc}(\mR^2 \setminus D)$, one may assume that $v_\eps \rightharpoonup v$ weakly in $H^1_{\loc}(\mR^2 \setminus D)$ (up to a subsequence). Then $v \in W^1(\mR^2 \setminus \bar D)$. This fact is not stated in \cite[Lemma 2.2]{NguyenHelmholtz} however the proof is already there (see \cite[(2.20) and (2.22)]{NguyenHelmholtz}). It is clear that $v$ satisfies \eqref{sysv2d}. Since \eqref{sysv2d} has a unique solution $v \in W^1(\mR^2 \setminus \bar D)$ (see e.g. \cite[Theorem 2.5.14 and the remarks on page 64]{Nedelec}) the conclusion holds for the whole sequence. The details of the proof are left to the reader. \proofend

\medskip
The following lemma plays an important role in the proof of Theorem~\ref{thm1-2d*}.

\begin{lemma}\label{lemd=2} Let $0 < \eps < 1$, $k> 0$,  $\theta_\eps \in L^2(D)$, $g_\eps \in H^{-\frac{1}{2}}(\partial D)$, $a$ be a real symmetric matrix-valued function, and $\sigma$ be a complex function defined on $D$. Assume that $a$ is uniformly elliptic, $0< \ess \inf \Re \sigma \le \ess \sup \Re \sigma < + \infty$ and $0 \le \ess \inf \Im \sigma \le \ess \sup \Im \sigma < + \infty$, and the system is non-resonant \footnote{The definition of the non-resonance in this case is the same as the case corresponding to the unit ball $B_1$ i.e. if $w \in W^1(\mR^2)$ satisfies \eqref{uniqueness2d} where $B_1$ is replaced by $D$, then $w =0$.}. Let $v_\eps \in H^1_{\loc}(\mR^2)$ be the unique solution of
\begin{equation*}
\left\{\begin{array}{ll}
\Delta v_\eps + \eps^2 k^2 v_\eps = 0 & \mbox{in } \mR^2 \setminus \overline D, \\[12pt]
\dive  (a \nabla v_\eps) + k^2 \sigma v_\eps = \theta_\eps & \mbox{in } D, \\[12pt]
\dsp \frac{\partial v_\eps}{\partial \eta}\Big|_{\mathrm{ext}} - a \nabla v_\eps \cdot \eta \Big|_{\mathrm{int}} = g_\eps & \mbox{on } \partial D \\[12pt]
v_\eps \mbox{ satisfies the outgoing condition}.
\end{array}\right.
\end{equation*}
Then

\begin{enumerate}
\item[$i)$] We have
\begin{equation*}
\|v_\eps \|_{H^1(B_5 \setminus D)} \le C ( \| \theta_\eps \|_{L^2(D)} +  \| g_\eps \|_{H^{-\frac{1}{2}}(\partial D)}),
\end{equation*}
for some positive constant $C$ depending only on $k$, $a$, $\sigma$ and $D$ but independent of $\eps$, $\theta_\eps$, and $g_\eps$.

\item[$ii)$] Assume in addition that  $\theta_\eps \rightharpoonup \theta$ weakly in $L^2(D)$ and $g_\eps \rightharpoonup g$ weakly in $H^{-\frac{1}{2}}(\partial D)$ as $\eps \to 0$. Then $v_\eps \rightharpoonup v$ weakly in $H^1(D)$. Consequently,
\begin{equation*}
\lim_{\eps \to 0}\|v_\eps - v \|_{L^2(D)} = 0.
\end{equation*}
Here $v \in W^1(\mR^2)$ is the unique solution of the system
\begin{equation}\label{sysv2d*}
\left\{\begin{array}{ll}
\Delta v= 0 & \mbox{in } \mR^2 \setminus \overline D, \\[12pt]
\dive (a \nabla v) + k^2 \sigma v = \theta & \mbox{in } D, \\[12pt]
\dsp \frac{\partial v}{\partial \eta}\Big|_{\mathrm{ext}} -   a \nabla v \cdot \eta  \Big|_{\mathrm{int}}  = g & \mbox{on } \partial D.
\end{array}\right.
\end{equation}
\end{enumerate}
\end{lemma}

\noindent{\bf Proof.} We first prove that
\begin{equation}\label{conclu1d=3-2}
\|v_\eps \|_{L^2(B_5)} \le C \big( \| \theta_\eps \|_{L^2(D)} + \| g_\eps \|_{H^{-\frac{1}{2}}(\partial D)} \big),
\end{equation}
by contradiction. Suppose that there exist $(\theta_n) \subset L^2(D)$, $(g_n) \subset H^{-\frac{1}{2}}(\partial D)$ and $(\eps_n)$ such that $(\eps_n)$ converges to $0$,
\begin{equation}\label{contradiction1-2}
\|v_n \|_{L^2(B_5)}  = 1 \quad \mbox{and} \quad \lim_{n \lr \infty} \big( \|\theta_n\|_{L^2(D)} + \| g_n \|_{H^{-\frac{1}{2}}(\partial D)} \big) = 0.
\end{equation}
Here $v_n \in H^1_{\loc}(\mR^2)$ is the unique solution of the system
\begin{equation}\label{sysvnd=3-2-1}
\left\{\begin{array}{ll}
\Delta v_n + \eps_n^2 k^2 v_n = 0 & \mbox{in } \mR^2 \setminus \overline D, \\[12pt]
\dive (a \nabla v_n) + k^2 \sigma v_n = \theta_n & \mbox{in } D, \\[12pt]
\dsp \frac{\partial v_n}{\partial \eta}\Big|_{\mathrm{ext}} -   a \nabla v_n \cdot \eta  \Big|_{\mathrm{int}} = g_n & \mbox{on } \partial D \\[12pt]
v_n \mbox{ satisfies the outgoing condition}.
\end{array}\right.
\end{equation}
Applying Lemma~\ref{lem0-2d} and the regularity theory of elliptic equations, we deduce from \eqref{contradiction1-2} that
\begin{equation}\label{estGradvnOut-2}
\|v_n \|_{H^1(B_r \setminus B_3)} \le C_r \quad \forall \, r > 5.
\end{equation}
Multiplying system~\eqref{sysvnd=3-2-1} with $\bar v_n$ (the conjugate of $v_n$) and integrating the expression obtained over $B_4$, we have
\begin{multline*}
\int_{B_4 \setminus D} |\nabla v_n|^2 - \eps_n^2 k^2 \int_{B_4 \setminus D} |v_n|^2 + \int_{D} \langle a\nabla v_n, \nabla v_n \rangle - \int_{D} k^2 \sigma |v_n|^2 \\[12pt]
= - \int_{D} \theta_n \bar v_n +  \int_{\partial B_4} \frac{\partial v_n}{\partial r} \bar v_n - \int_{\partial D} g_n \bar v_n.
\end{multline*}
From  \eqref{contradiction1-2} and \eqref{estGradvnOut-2}, it follows that
\begin{equation}\label{estGradvnInt-2}
\int_{B_4} |\nabla v_n|^2  \le C.
\end{equation}
Combining \eqref{contradiction1-2}, \eqref{estGradvnOut-2}, and \eqref{estGradvnInt-2} yields
\begin{equation}\label{estGradvn-2}
\|v_n \|_{H^1(B_r)} \le C_r \quad \forall \, r > 0.
\end{equation}
Hence without loss of generality, one may assume that $v_n \rightharpoonup v$ weakly in $H^1_{\loc}(\mR^2)$ and $v_n \to v$ in $L^2_{\loc}(\mR^2)$, where $v \in W^1(\mR^2)$ (by Lemma~\ref{lem0-2d}) is a solution of the system
\begin{equation*}
\left\{\begin{array}{ll}
\Delta v= 0 & \mbox{in } \mR^2 \setminus \bar D, \\[12pt]
\dive (a \nabla v) + k^2 \sigma v = 0 & \mbox{in } D, \\[12pt]
\dsp \frac{\partial v}{\partial \eta}\Big|_{\mathrm{ext}} -   a \nabla v \cdot \eta  \Big|_{\mathrm{int}} = 0 & \mbox{on } \partial D.
\end{array}\right.
\end{equation*}
Since the system is non-resonant, $v =0$. We have a contradiction since
$$\| v\|_{L^2(B_5)} = \lim_{n \to \infty}\|v_n \|_{L^2(B_5)} = 1.$$
Hence \eqref{conclu1d=3-2} is proved. Using the same argument to obtain \eqref{estGradvn-2}, we deduce part $i)$ of the conclusion from \eqref{conclu1d=3-2}. Part $ii)$ of the conclusion follows from part $i)$ and the uniqueness of system~\eqref{sysv2d*}. \proofend

\subsection{\bf Proof of Theorem~\ref{thm1-2d*}.} The proof of Theorem~\ref{thm1-2d*} is quite similar to the one of Theorem~\ref{thm1-3d*}. However instead of using Lemmas~\ref{lem0-3d}, \ref{lemUniqueness-3d}, and \ref{lemd=3}, we apply Lemmas~\ref{lem0-2d} and \ref{lemd=2}. For the convenience of the reader, we present the proof.

Let $u_{1, \eps} \in H^1_{\loc}(\mR^2)$ be the unique solution of
\begin{equation*}
\left\{\begin{array}{ll}
\Delta u_{1, \eps} + k^2 u_{1, \eps} = f ~(= {F_\eps^{-1}}_{*}f) & \mbox{ in } \mR^2 \setminus \overline B_\eps,\\[12pt]
u_{1, \eps} = 0 & \mbox{ in } B_\eps \\[12pt]
u_{1, \eps} \mbox{ satisfies the outgoing condition}. &
\end{array}\right.
\end{equation*}
Define
\begin{equation*}
w_{1, \eps} = u_{1, \eps} - u,
\end{equation*}
and
\begin{equation*}
w_{2, \eps} = u_\eps - u_{1, \eps}.
\end{equation*}

We claim that
\begin{equation}\label{claim-w1-2d}
\|w_{1, \eps} \|_{H^1(B_{4r} \setminus B_{r})} \le \frac{C_r}{|\ln \eps |} \| f \|_{L^2}
\end{equation}
and
\begin{equation}\label{claim-w2-2d}
 \|w_{2, \eps} \|_{H^1(B_{4r} \setminus B_{r})} \le \frac{C_r}{|\ln \eps |} \| f \|_{L^2}.
\end{equation}

\medskip
\noindent\underline{Step 1:} Proof of Claim~\eqref{claim-w1-2d}. From the definition of $w_{1, \eps}$, it follows that $w_{1, \eps} \in H^1_{\loc}(\mR^2)$ and $w_{1, \eps}$ satisfies
\begin{equation*}
\left\{\begin{array}{ll}
\Delta w_{1, \eps} + k^2 w_{1, \eps} = 0 & \mbox{ in } \mR^2 \setminus \overline B_\eps,\\[12pt]
w_{1, \eps} = - u & \mbox{ in } \partial B_\eps \\[12pt]
w_{1, \eps} \mbox{ satisfies the outgoing condition}. &
\end{array}\right.
\end{equation*}
Define $W_{1, \eps}(x) = w_{1, \eps}(\eps x)$. Then $W_{1, \eps} \in H^1_{\loc}(\mR^2)$ and $W_{1, \eps}$ satisfies
\begin{equation}\label{defW1eps-2d}
\left\{\begin{array}{ll}
\Delta W_{1, \eps} + \eps^2 k^2 W_{1, \eps} = 0 & \mbox{ in } \mR^2 \setminus \overline B_1,\\[12pt]
W_{1, \eps} = -  u (\eps \cdot) & \mbox{ in } \partial B_1 \\[12pt]
W_{1, \eps} \mbox{ satisfies the outgoing condition}. &
\end{array}\right.
\end{equation}
Since $\| u (\eps \cdot)\|_{H^\frac{1}{2} (\partial B_1)} \le \| f\|_{L^2}$, by Lemma~\ref{lem0-2d}, we have
\begin{equation*}
\eps^2 \int_{B_{4r/ \eps} \setminus B_{r/ \eps}} |W_{1, \eps}|^2 \le \frac{C_r}{|\ln \eps|^2} \| f\|_{L^2}^2.
\end{equation*}
By a change of variables,
\begin{equation}\label{reg1-2d}
\int_{B_{4r} \setminus B_{r}} |w_{1, \eps}|^2 \le \frac{C_r}{|\ln \eps|^2} \| f\|_{L^2}^2.
\end{equation}
Since $\Delta w_{1, \eps} + k^2 w_{1, \eps} = 0$ in $\mR^2 \setminus \bar B_{\eps}$, by the regularity theory of elliptic equations,  Claim~\eqref{claim-w1-2d} follows from \eqref{reg1-2d}.

 \medskip
\noindent\underline{Step 2:} Proof of Claim~\eqref{claim-w2-2d}. It is clear that $w_{2, \eps} \in H^1_{\loc}(\mR^2)$ is the unique solution of
\begin{equation*}
\left\{\begin{array}{ll}
\Delta w_{2, \eps}  +  k^2 w_{2, \eps} = 0 & \mbox{ in } \mR^2 \setminus \overline B_\eps, \\[12pt]
\dive  (A_\eps \nabla w_{2, \eps}) + k^2 \Sigma_\eps w_{2, \eps} = {F_\eps^{-1}}_{*} f & \mbox{ in } B_\eps, \\[12pt]
\dsp \frac{\partial w_{2, \eps}}{\partial \eta} \Big|_{\mathrm{ext}} -  A_\eps \nabla w_{2, \eps} \cdot \eta \Big|_{\mathrm{int}}= - \frac{\partial u_{1, \eps}}{\partial \eta} & \mbox{ on } \partial B_\eps, \\[12pt]
w_{2, \eps} \mbox{ satisfies the outgoing condition.} &
\end{array}\right.
\end{equation*}
Define $W_{2, \eps}(x) = w_{2, \eps}(\eps x)$. Then $W_{2, \eps} \in H^1_{\loc}(\mR^2)$ and $W_{2, \eps}$ is the unique solution of
\begin{equation*}
\left\{\begin{array}{ll}
\Delta W_{2, \eps} + \eps^2 k^2 W_{2, \eps} = 0 & \mbox{in } \mR^2 \setminus \overline B_1, \\[12pt]
\dive (a \nabla W_{2, \eps}) + k^2 \sigma W_{2, \eps} = f & \mbox{in } B_1, \\[12pt]
\dsp \frac{\partial W_{2, \eps}}{\partial \eta}\Big|_{\mathrm{ext}} - a \nabla  W_{2, \eps} \cdot \eta \Big|_{\mathrm{int}} = - \eps \frac{\partial u_{1, \eps}}{\partial \eta} (\eps x) & \mbox{on } \partial B_1, \\[12pt]
W_{2, \eps} \mbox{ satisfies the outgoing condition}. &
\end{array}\right.
\end{equation*}
Since  $u_{1, \eps} = w_{1, \eps} + u$ and $W_{1, \eps} = w_{1, \eps} (\eps \cdot)$, it follows that
\begin{equation}\label{tt2}
\eps \frac{\partial u_{1, \eps}}{\partial \eta} (\eps x) = \frac{\partial W_{1, \eps}}{\partial \eta}(x) + \eps \frac{\partial u}{\partial \eta} (\eps x) \quad \mbox{ on } \partial B_1.
\end{equation}
From \eqref{defW1eps-2d} and Lemma~\ref{lem0-2d}, we obtain
\begin{equation*}
\Big\|\eps \frac{\partial u_{1, \eps}}{\partial \eta} (\eps x) \Big\|_{H^{-\frac{1}{2}}(\partial B_1)} \le C \| f\|_{L^2}.
\end{equation*}
We deduce from Lemma~\ref{lemd=2} that
\begin{equation*}
\| W_{2, \eps}\|_{H^1(B_5)} \le C \| f\|_{L^2}.
\end{equation*}
Applying Lemma~\ref{lem0-2d}, we have
\begin{equation*}
\eps^2 \int_{B_{4r/\eps} \setminus B_{r/\eps}} |W_{2, \eps}|^2 \le \frac{C_r}{|\ln \eps|^2} \| f\|_{L^2}^2.
\end{equation*}
By a change of variables,
\begin{equation*}
\int_{B_{4r} \setminus B_{r}} |w_{2, \eps}|^2 \le \frac{C_r}{|\ln \eps|^2}  \| f\|_{L^2}^2.
\end{equation*}
Claim~\eqref{claim-w2-2d} now follows from the regularity theory of elliptic equations and the fact that $\Delta w_{2, \eps} + k^2 w_{2, \eps} = 0$ in $\mR^2 \setminus \bar B_\eps$.

\medskip
Thus Claims~\eqref{claim-w1-2d} and \eqref{claim-w2-2d} are proved.  Since $u_\eps - u = w_{1, \eps} + w_{2, \eps}$, \eqref{est-2d} follows.

\medskip

The rest of the proof now follows from the following facts:
\begin{enumerate}
\item[$a)$] $W_{1, \eps} \rightharpoonup W_1 = - u(0)$ weakly in $H^1_{\loc}(\mR^2 \setminus \overline B_1)$ since  $W_1 \in W^1(\mR^2 \setminus \overline B_1)$ is the unique solution of
\begin{equation}\label{sysW}
\left\{\begin{array}{ll}
\Delta W_1 = 0 & \mbox{ in } \mR^2 \setminus \overline B_1,\\[12pt]
W_1 = -  u (0) & \mbox{ in } \partial B_1.
\end{array}\right.
\end{equation}
by Lemma~\ref{lem0-2d}.

\item[$b)$] $W_{2, \eps} \rightharpoonup v$ weakly in $W^1_{\loc}(\mR^2)$ where $ v$  is the unique solution of the system
    \begin{equation*}\left\{
\begin{array}{ll}
\Delta v  = 0 & \mbox{ in } \mR^2 \setminus \overline B_1, \\[12pt]
\dive(a \nabla   v) + k^2 \sigma  v = f & \mbox{ in } B_1, \\[12pt]
\dsp \frac{\partial v}{\partial \eta} \Big|_{\mathrm{ext}} = a \nabla  v \cdot \eta & \mbox{ on } \partial B_1,
\end{array}\right.
\end{equation*}
since $W_1 = - u(0)$ in $\mR^2 \setminus B_1$ by statement a), and \eqref{tt2} holds.

\item[$c)$] $U_\eps = W_{2, \eps}$ in $B_1$ and $v =Cl_2(f)$.
\end{enumerate}
\proofend

\subsection{Proof of Proposition~\ref{proNonRes}}

Proposition~\ref{proNonRes} is a consequence of the following lemma.

\begin{lemma}\label{lemUniqueness-2d} Let $k > 0$, $0< c_1 < c_2 < \infty$, $\theta \in L^2(B_1)$ and $g \in H^{\frac{1}{2}}(\partial B_1)$. Assume that $c_1 |\xi|^2 \le \langle a \xi,  \xi \rangle \le c_2 |\xi|^2$, $0 \le \Im (\sigma) \le c_2$, $c_1 < \Re (\sigma) < c_2$ on $B_1$. Then there exists $k_0>0$ depending only on $c_1$, and $c_2$ such that if $0 < k < k_0$ then there exists  a unique solution $v\in W^1(\mR^2)$ of the system
\begin{equation}\label{sysvnd=3-2}
\left\{\begin{array}{ll}
\Delta v= 0 & \mbox{in } \mR^2 \setminus \overline B_1, \\[12pt]
\dive (a \nabla v) + k^2 \sigma v = \theta & \mbox{in } B_1, \\[12pt]
\dsp \frac{\partial v}{\partial \eta}\Big|_{\mathrm{ext}} =   a \nabla v \cdot \eta  \Big|_{\mathrm{int}} + g & \mbox{on } \partial B_1.
\end{array}\right.
\end{equation}
\end{lemma}

\noindent {\bf Proof.} We only prove the uniqueness of $v$. The existence of $v$ follows from the uniqueness. For this end, it suffices to prove that $v = 0$ if $v \in W^1(\mR^2)$ is a solution of
\begin{equation}\label{sysvnd=3-2}
\left\{\begin{array}{ll}
\Delta v= 0 & \mbox{in } \mR^2 \setminus \overline B_1, \\[12pt]
\dive (a \nabla v) + k^2 \sigma v = 0 & \mbox{in } B_1, \\[12pt]
\dsp \frac{\partial v}{\partial \eta}\Big|_{\mathrm{ext}} =   a \nabla v \cdot \eta  \Big|_{\mathrm{int}} & \mbox{on } \partial B_1.
\end{array}\right.
\end{equation}
Since $v \in W^1(\mR^2)$, it follows from \eqref{sysvnd=3-2} that (see e.g. \cite{Nedelec} or \cite[Proposition 2]{NguyenVogelius})
\begin{equation}\label{null-cond}
\int_{\partial B_1} \frac{\partial v}{\partial \eta}\Big|_{\mathrm{ext}} = 0 \; \footnote{This identity is the key of the proof.}
\end{equation}
and
\begin{equation}\label{iden1}
\int_{\mR^2 \setminus B_1} |\nabla v|^2 + \int_{B_1} \langle a \nabla v, \nabla v \rangle - \int_{B_1} k^2 \sigma |v|^2 =0.
\end{equation}
Integrating the equation $\dive (a \nabla v) + k^2 \sigma v = 0$ on $B_1$ and using \eqref{null-cond}, we have
\begin{equation}\label{null-cond-1}
\int_{B_1} \sigma v = 0,
\end{equation}
which implies, by a standard compactness argument,
\begin{equation}\label{ineq2}
\|\nabla v \|_{L^2(B_1)}  \ge C \| v\|_{L^2(B_1)},
\end{equation}
for some positive constant $C$ depending only on $c_1$ and $c_2$. Combining \eqref{iden1} and \eqref{ineq2} yields
\begin{equation*}
\int_{\mR^2 \setminus B_1} |\nabla v|^2 + \int_{B_1} \langle a \nabla v, \nabla v \rangle = 0
\end{equation*}
if $0 < k < k_0$ for some $k_0 > 0$. This implies that $v$ is constant. Hence $v =0$ by \eqref{null-cond-1}.
\proofend

\section{Proof of Proposition~\ref{proInfinite}}\label{SectionInfinite}

Define $U_\eps = u_\eps(\eps x)$ where $u_\eps = u_c \circ F_\eps$. Then $U_\eps$ is the solution of
\begin{equation*}
\left\{\begin{array}{ll}
\Delta U_\eps + \eps^2 k^2 U_\eps = 0 & \mbox{in } \mR^3 \setminus \overline B_1, \\[12pt]
\dive (a \nabla U_\eps) + k^2 \sigma U_\eps = \be & \mbox{in } B_1, \\[12pt]
\dsp \frac{\partial U_\eps}{\partial \eta}\Big|_{\mathrm{ext}} - \frac{1}{\eps} a \nabla U_\eps \cdot \eta \Big|_{\mathrm{int}} = 0 & \mbox{on } \partial B_1, \\[12pt]
U_\eps \mbox{ satisfies the outgoing condition}. &
\end{array}\right.
\end{equation*}

\medskip
\noindent \underline{Step 1:} Proof of part $i)$. It suffices to prove
\begin{equation}\label{Energy-2}
\liminf_{\eps \to 0} \eps \| U_\eps\|_{H^\frac{1}{2}( \partial B_1)} > 0,
\end{equation}
since $u_c = U_\eps$ in $B_1$.  We will prove \eqref{Energy-2} by contradiction. Suppose that there exists $\eps_n \to 0$ such that
\begin{equation*}
\lim_{n \to \infty} \eps_n \| U_n\|_{H^\frac{1}{2}(\partial B_1)} =0,
\end{equation*}
where $U_n$ is the solution of
\begin{equation*}
\left\{\begin{array}{ll}
\Delta U_n + \eps_n^2 k^2 U_n = 0 & \mbox{in } \mR^3 \setminus \overline B_1, \\[12pt]
\dive (a \nabla U_n) + k^2 \sigma U_n = \be & \mbox{in } B_1, \\[12pt]
\dsp \frac{\partial U_n}{\partial \eta}\Big|_{\mathrm{ext}} - \frac{1}{\eps_n} a \nabla U_n \cdot \eta \Big|_{\mathrm{int}} = 0 & \mbox{on } \partial B_1, \\[12pt]
U_n \mbox{ satisfies the outgoing condition}. &
\end{array}\right.
\end{equation*}
By Lemma~\ref{lem0-3d}, we have
\begin{equation*}
\lim_{n \to \infty} \eps_n \| U_n \|_{H^1(B_r \setminus \overline B_1)}  = 0 \quad \forall \, r > 1,
\end{equation*}
which implies, since $\Delta U_n + \eps_n^2 k^2 U_n = 0$ in $\mR^3 \setminus \overline B_1$,
\begin{equation*}
\lim_{n \to \infty} \eps_n \Big\| \frac{\partial U_n}{\partial \eta} \Big|_{\mathrm{ext}} \Big\|_{H^{-\frac{1}{2}}(\partial B_1)} =0.
\end{equation*}
Hence, since $\dsp \frac{\partial U_n}{\partial \eta}\Big|_{\mathrm{ext}} - \frac{1}{\eps_n} a \nabla U_n \cdot \eta \Big|_{\mathrm{int}} = 0$  on $\partial B_1$, we have
\begin{equation}\label{trace2}
\lim_{n \to \infty} \Big \| a \nabla U_n \cdot \eta \Big|_{\mathrm{int}} \Big\|_{H^{-\frac{1}{2}}(\partial B_1)} = 0.
\end{equation}
Multiplying the equation $\dive (a \nabla U_n) + k^2 \sigma U_n = \be$ by $\bar \be$, integrating the expression obtained on $B_1$, and using the fact that $\be \in M$, we have
\begin{equation}\label{trace3}
\int_{\partial B_1} \big(a \nabla U_n \cdot \eta \big) \bar \be = \int_{B_1} |\be|^2 =1.
\end{equation}
Combining \eqref{trace2} and \eqref{trace3}, we have a contradiction.

\medskip
\noindent \underline{Step 2:} Proof of part $ii)$. Since $\be$ is radial, $a = \lambda_1 I$, and $\sigma = \lambda_2$ for some positive constants $\lambda_1$ and $\lambda_2$, $U_\eps$ is radial. This implies $\dsp U_\eps (x) = c_\eps \frac{e^{i k \eps |x| }}{|x|}$ if $|x| > 1$ for some positive constant $c_\eps$. Hence the conclusion of part $ii)$ follows from \eqref{Energy-2} and the fact that $u_c(x) = U_\eps(x/ \eps)$ for $x \in B_4 \setminus B_2$. \proofend

\section{Proof of Proposition~\ref{proInfinite-2d}}\label{SectionInfinite-2d}

Define $U_\eps = u_\eps(\eps x)$ where $u_\eps = u_c \circ F_\eps$. Then $U_\eps$ is the solution of
\begin{equation*}
\left\{\begin{array}{ll}
\Delta U_\eps + \eps^2 k^2 U_\eps = 0 & \mbox{in } \mR^2 \setminus \overline B_1, \\[12pt]
\dive (a \nabla U_\eps) + k^2 \sigma U_\eps = \be & \mbox{in } B_1, \\[12pt]
\dsp \frac{\partial U_\eps}{\partial \eta}\Big|_{\mathrm{ext}} - a \nabla U_\eps \cdot \eta \Big|_{\mathrm{int}} = 0 & \mbox{on } \partial B_1, \\[12pt]
U_\eps \mbox{ satisfies the outgoing condition}. &
\end{array}\right.
\end{equation*}

\medskip
\noindent \underline{Step 1:} Proof of part $i)$. It suffices to prove
\begin{equation}\label{Energy-2-2d}
\liminf_{\eps \to 0}  \| U_\eps\|_{H^1(B_1)} = \infty,
\end{equation}
since $u_c = U_\eps$ in $B_1$.  We will prove \eqref{Energy-2-2d} by contradiction. Suppose that there exists $\eps_n \to 0$ such that
\begin{equation}\label{energy1}
\sup_{n \in \mN}  \| U_n\|_{H^1(B_1)} < + \infty,
\end{equation}
where $U_n$ is the solution of
\begin{equation*}
\left\{\begin{array}{ll}
\Delta U_n + \eps_n^2 k^2 U_n = 0 & \mbox{in } \mR^2 \setminus \overline B_1, \\[12pt]
\dive (a \nabla U_n) + k^2 \sigma U_n = \be & \mbox{in } B_1, \\[12pt]
\dsp \frac{\partial U_n}{\partial \eta}\Big|_{\mathrm{ext}} -  a \nabla U_n \cdot \eta \Big|_{\mathrm{int}} = 0 & \mbox{on } \partial B_1, \\[12pt]
U_n \mbox{ satisfies the outgoing condition}. &
\end{array}\right.
\end{equation*}
By Lemma~\ref{lem0-2d}, we have
\begin{equation*}
\sup_{n \in \mN} \| U_n \|_{H^1(B_r \setminus \overline B_1)}  < + \infty \quad \forall \, r > 1.
\end{equation*}
It follows from \eqref{energy1} that
\begin{equation*}
\sup_{n \in \mN} \| U_n \|_{H^1(B_r)}  < + \infty \quad \forall \, r > 1.
\end{equation*}
Applying Lemma~\ref{lem0-2d}, we may assume that $U_n \rightharpoonup U$ in $H^1_{\loc}(\mR^2)$ for some $U \in W^1(\mR^2)$ which satisfies
\begin{equation}\label{contra1}
\left\{\begin{array}{ll}
\Delta U  = 0 & \mbox{in } \mR^2 \setminus \overline B_1, \\[12pt]
\dive (a \nabla U) + k^2 \sigma U = \be & \mbox{in } B_1, \\[12pt]
\dsp \frac{\partial U}{\partial \eta}\Big|_{\mathrm{ext}} -  a \nabla U \cdot \eta \Big|_{\mathrm{int}} = 0 & \mbox{on } \partial B_1, \\[12pt]
\end{array}\right.
\end{equation}
However, system~\eqref{contra1} has no solution in $W^1(\mR^2)$ since $\be \in N$. We have a contradiction.

\medskip
\noindent \underline{Step 2:} Proof of part $ii)$. For notational ease, we will assume that $\lambda_1 =  1$ and $\lambda_2 = \lambda$.
Since $\be$ is radial, $a = I$, and $\sigma = \lambda$, $U_\eps$ is radial. Thus $U_\eps$ can be written under the form
\begin{equation*}
U_\eps = c_\eps H^{(1)}_0(\eps k r) \quad \mbox{ for } r = |x| \ge 1.
\end{equation*}
This implies
\begin{equation}\label{der0}
\frac{\partial U_\eps}{\partial \eta} = c_\eps \eps k \partial_r H^{(1)}_0(\eps k) \quad \mbox{ for } r = |x| = 1.
\end{equation}
Since $\be$ is radial, $\be \in W^1(\mR^2 \setminus \overline B_1)$, and $\Delta \be = 0$ in $\mR^2 \setminus \overline B_1$, it follows that  $\be$ is constant in $\mR^2 \setminus B_1$. Hence
\begin{equation}\label{der}
\frac{\partial \be }{\partial \eta} = 0 \quad \mbox{ on } \partial B_1.
\end{equation}
It is clear that
\begin{equation*}
\Delta U_\eps + k^2 \lambda U_\eps = \be \mbox{ in } B_1  \quad \mbox{ and } \quad \Delta \be  + k^2 \lambda  \be  = 0 \mbox{ in } B_1.
\end{equation*}
Multiplying the first equation by $\bar \be$, the second equation by $\bar U_\eps$ and integrating the obtained expressions on $B_1$, we have
\begin{equation*}
\int_{B_1} \nabla U_\eps \nabla \bar \be  -  k^2 \lambda U_\eps \bar \be =  - \int_{B_1} |\be|^2 + \int_{\partial B_1} \frac{\partial U_\eps}{\partial \eta} \bar \be,
\end{equation*}
and
\begin{equation*}
\int_{B_1} \nabla \be \nabla \bar U_\eps  - k^2  \lambda \be \bar U_\eps = \int_{\partial B_1} \frac{\partial \be}{\partial \eta} \bar U_\eps =0.
\end{equation*}
(in the last equality, we used \eqref{der}).  This implies
\begin{equation*}
1= \int_{B_1} |\be|^2 =  \int_{\partial B_1} \frac{\partial U_\eps}{\partial \eta} \bar \be.
\end{equation*}
It follows from \eqref{der0} that
\begin{equation*}
\int_{\partial B_1} \bar \be c_\eps \eps k \partial_r  H^{(1)}_0 (\eps k) = 1.
\end{equation*}
We recall that, see e.g., \cite[Chapter 3]{ColtonKressInverse},
\begin{equation}\label{Hankel0}
\lim_{r \lr 0} \frac{r dH_{0}^{(1)}(r)}{dr} = - \frac{2}{ i \pi},
\end{equation}
which yields
\begin{equation*}
|c_\eps| \ge c,
\end{equation*}
for some positive constant $c$. Hence
\begin{equation*}
|U_\eps (x/ \eps)| \ge c |H^{(1)}_0(k |x|)| \quad \mbox{ for }  2 \le |x| \le 4.
\end{equation*}
Since $u_c(x) = U_\eps(x/ \eps)$ for $x \in B_4 \setminus B_2$, the conclusion of part $ii)$ follows. \proofend

\section{Proof of Proposition~\ref{proinstability}}\label{SectionInstability}

This section is devoted to the proof of Proposition~\ref{proinstability}. It suffices to prove the following proposition.

\begin{proposition} Let $d=2,3$, $k> 0$, and $\sigma_0 > 0$ be such that $j_0'(k \sigma_0) =0$ if $d=3$ and $J_0'(k \sigma_0) =0$ if $d=2$. Then there exists a family of positive numbers  $(\sigma_\eps)$ such that
\begin{equation*}\left\{
\begin{array}{cl}
0 < \liminf_{\eps \to 0} \eps^{-1} |\sigma_\eps - \sigma_0| \le \limsup_{\eps \to 0} \eps^{-1} |\sigma_\eps - \sigma_0| < + \infty & \mbox{ if } d= 3, \\[12pt]
0 < \liminf_{\eps \to 0} |\ln \eps| |\sigma_\eps - \sigma_0| \le \limsup_{\eps \to 0} |\ln \eps| |\sigma_\eps - \sigma_0| < + \infty & \mbox{ if } d= 2,
\end{array} \right.
\end{equation*}
and
\begin{equation*}
\| u_{c, s} \|_{L^2(B_4 \setminus B_2)} \ge \left\{ \begin{array}{cl} \| h_0 \|_{L^2(B_4 \setminus B_2)} & \mbox{ if } d =3, \\[6pt]
\| H_0 \|_{L^2(B_4 \setminus B_2)} & \mbox{ if } d =2.
\end{array}\right.
\end{equation*}
Here $u_{c, s} \in H^1_{\loc}(\mR^d)$ is such that $u_{c, s}$ satisfies the outgoing condition and if $u_{c}:  = u_{c, s} + u_{c, i}$ with
\begin{equation*}
u_{c, i} = \left\{ \begin{array}{cl}  j_0(k |x|) & \mbox{ if } d =3, \\[6pt]
J_0(k |x|) & \mbox{ if } d =2,
\end{array}\right.
\end{equation*}
then $u_c$ is a solution of the equation
\begin{equation*}
\dive(A_c \nabla u_c) + k^2 \Sigma_c u_c = 0,
\end{equation*}
where $(A_c, \Sigma_c)$ is defined in \eqref{defASigmac} with $a =I$ and $\sigma = \sigma_\eps$.
\end{proposition}

Hereafter $h_0$ denotes the spherical Hankel function of the first kind of order $0$, $j_0 = \Re(h_0)$,  and $y_0 = \Im(h_0)$, and $H_0$ denotes the  Hankel function of the first kind of order $0$, $J_0 = \Re(H_0)$, and $Y_0 = \Im(H_0)$.

\medskip

\noindent{\bf Proof.}  Set $u_\eps = u_{c} \circ F_\eps$ and $u_{\eps, s} = u_\eps - u_{c, i}$ where $F_\eps$ is given in \eqref{defFeps}. According to Proposition~\ref{fundPro}, $u_\eps = u_c$ and $u_{\eps, s} = u_{c, s}$ in $\mR^d \setminus B_2$, $u_{\eps, s}$ satisfies the outgoing condition, and $u_\eps$ is the solution of the equation
\begin{equation*}
\dive(A_\eps \nabla u_\eps) + k^2 \Sigma_\eps u_\eps = 0,
\end{equation*}
where
\begin{equation*}
A_\eps, \Sigma_\eps =
\left\{\begin{array}{cl}
I, 1  & \mbox{if } x \in \mR^d \setminus B_{\eps}, \\[12pt]
\dsp \frac{I}{\eps^{d-2}}, \frac{\sigma_\eps}{\eps^d}  & \mbox{if } x \in B_{\eps}.
\end{array}\right.
\end{equation*}
Define $U_\eps(x) = u_\eps(\eps x)$ and $U_{\eps, s}(x) = u_{\eps, s}(\eps x) = U_\eps(x) - u_{c, i}(\eps x)$ for $x \in \mR^d$. Then $U_{\eps, s} (x)= u_{c, s}(\eps x)$ for $|x|> 2/ \eps$, $U_\eps$ satisfies the equation
\begin{equation*}
\dive(\tilde A_\eps \nabla U_\eps) +  k^2 \tilde \Sigma_\eps U_\eps = 0,
\end{equation*}
and $U_{\eps, s}$ satisfies the outgoing condition.  Here
\begin{equation*}
\tilde A_\eps, \tilde \Sigma_\eps =
\left\{\begin{array}{cl}
I, \eps^2  & \mbox{if } x \in \mR^d \setminus B_1, \\[12pt]
\dsp \frac{I}{\eps^{d-2}}, \frac{\sigma_\eps}{\eps^{d-2}} \ & \mbox{if } x \in B_1.
\end{array}\right.
\end{equation*}

\medskip
\noindent \underline{Step 1:} $d=3$. It is clear that
\begin{equation*}
U_{\eps, i} = j_0( k \eps |x|), \quad  U_{\eps, s} = \alpha_0 h_0( k \eps |x|) \quad \mbox{ for } |x| > 1,
\end{equation*}
and
\begin{equation*}
U_{\eps, t} =  \beta_0 j_0(k_\eps |x|)  \quad \mbox{ for } |x| < 1,
\end{equation*}
where $k_\eps = k \sigma_\eps$, $U_{\eps, t} := U_{\eps, s} + U_{\eps, i} = U_{\eps}$ in $B_1$, for some $\alpha_0, \beta_0$. Using the transmission conditions, namely,
\begin{equation*}
\left\{\begin{array}{ll}
U_{\eps, s} + U_{\eps, i} = U_{\eps, t} & \mbox{ on } \partial B_1, \\[12pt]
\dsp \frac{\partial U_{\eps, s}}{\partial r} + \frac{\partial U_{\eps, i}}{\partial r} = \frac{1}{\eps} \frac{\partial U_{\eps, t}}{\partial r} & \mbox{ on } \partial B_1,
\end{array}\right.
\end{equation*}
we have
\begin{equation*}
\left\{\begin{array}{ll}
\alpha_0 h_0(k \eps)  +  j_0(k \eps)  = \beta_0 j_0(k_\eps) & \mbox{ on } \partial B_1, \\[12pt]
\alpha_0 k  \eps h_0'(\eps)  +  k \eps j_0'(\eps)  = \frac{1}{\eps}  k_\eps \beta_0 j_0'(k_\eps) & \mbox{ on } \partial B_1.
\end{array}\right.
\end{equation*}
It follows that
\begin{equation}\label{definition-alpha0}
\alpha_0 =  - \frac{k \eps j_0'(k \eps) j_0(k_\eps) - \frac{1}{\eps} k_\eps j_0(k \eps) j_0'(k_\eps)}{k \eps h_0'(k \eps) j_0(k_\eps) - \frac{1}{\eps} k_\eps h_0(k \eps) j_0'(k_\eps)}.
\end{equation}
Since $y_0(t) = \cos t / t$, we have
\begin{equation*}
k \eps^2 \frac{y_0'(k \eps)}{y_0(k \eps)} = k \eps^2 \Big(- \frac{1}{k \eps} - \frac{\sin (k \eps)}{\cos (k  \eps)} \Big) = - \eps - k \eps^2 \sin (k \eps)/ \cos (k \eps).
\end{equation*}
Let $\sigma_\eps $ be such that $k_\eps = k \sigma_\eps$ converges to $k \sigma_0$ and
\begin{equation*}
\frac{j_0'(k_\eps)}{j_0(k_\eps)}= - \eps - k \eps^2 \sin (k \eps)/ \cos ( k \eps).
\end{equation*}
Then the complex part in the denominator of the RHS of \eqref{definition-alpha0} equals 0. On the other hand, the real part in the denominator of the RHS of \eqref{definition-alpha0} equals the numerator of the RHS of \eqref{definition-alpha0}. Hence it follows from \eqref{definition-alpha0} that
\begin{equation*}
\alpha_0 = -1,
\end{equation*}
which implies the conclusion in the $3d$ case since $u_{c, s}(x) = U_{\eps, s}(x/ \eps)$ for $|x| > 2$.

\medskip

\noindent \underline{Step 2:} $d=2$. It is clear that
\begin{equation*}
U_{\eps, i} = J_0( k \eps |x|), \quad  U_{\eps, s} = \alpha_0 H_0( k \eps |x|) \quad \mbox{ for } |x| > 1,
\end{equation*}
and
\begin{equation*}
U_{\eps, t} =  \beta_0 J_0(k_\eps |x|)  \quad \mbox{ for } |x| < 1,
\end{equation*}
where $k_\eps = k \sigma_\eps$, $U_{\eps, t} := U_{\eps, s} + U_{\eps, i} = U_{\eps}$ in $B_1$, for some $\alpha_0, \beta_0$.  Using the transmission conditions, namely,
\begin{equation*}
\left\{\begin{array}{ll}
U_{\eps, s} + U_{\eps, i} = U_{\eps, t} & \mbox{ on } \partial B_1, \\[12pt]
\dsp \frac{\partial U_{\eps, s}}{\partial r} + \frac{\partial U_{\eps, i}}{\partial r} = \frac{\partial U_{\eps, t}}{\partial r} & \mbox{ on } \partial B_1,
\end{array}\right.
\end{equation*}
we have
\begin{equation*}
\left\{\begin{array}{ll}
\alpha_0 H_0(k \eps)  +  J_0(k \eps)  = \beta_0 J_0(k_\eps) & \mbox{ on } \partial B_1, \\[12pt]
\alpha_0 k \eps H_0'(k \eps)  +  k \eps J_0'(k \eps)  =  k_\eps \beta_0 J_0'(k_\eps) & \mbox{ on } \partial B_1,
\end{array}\right.`
\end{equation*}
Thus it follows that
\begin{equation}\label{definition-alpha2}
\alpha_0 =  - \frac{k \eps J_0'(k \eps) J_0(k_\eps) -  k_\eps J_0(k \eps) J_0'(k_\eps)}{k \eps H_0'(k \eps) J_0(k_\eps) -  k_\eps H_0(k \eps) J_0'(k_\eps)}.
\end{equation}
Since $Y_0(t) =  \frac{2}{\pi} \ln (t/2)$, we have
\begin{equation*}
k \eps \frac{Y_0'(k \eps)}{Y_0(k \eps)} = \frac{1}{\ln (k \eps/2)}.
\end{equation*}
Let $\sigma_\eps$ be such that $k_\eps = k \sigma_\eps \to k \sigma_0$ and
\begin{equation*}`
\frac{k_\eps J_0'(k_\eps)}{J_0(k_\eps)} = \frac{1}{\ln (k \eps/2)}.
\end{equation*}
Then as in the $3d$ case, $\alpha_0 = -1$, and the conclusion in the $2d$ case follows from the fact that $u_{c, s}(x) = U_{\eps, s}(x/ \eps)$ for $|x| > 2$.
\proofend

%\begin{remark} In \cite{RuanYanNeffQiu}, the authors argued to obtain the degree of invisibility in $2d$. It seems for us that their arguments works for any finite range of $\sigma$. We wonder how their arguments are compatible with our instability result in $2d$.
%\end{remark}

\medskip
\noindent{\bf Acknowledgments.} The author thanks Bob Kohn for stimulating discussions and useful comments on the manuscript.

\input{No-lossy10-extension.bbl}

%\bibliographystyle{amsplain}
%\bibliography{bib1}

\end{document}

%% file: macros.tex
\newtheorem{theorem}{Theorem}
\newtheorem{lemma}{Lemma}

\newtheorem{proposition}{Proposition}

\newtheorem{remark}{Remark}

\newtheorem{definition}{Definition}
\numberwithin{equation}{section}

\newcommand{\proofend}{\hfill $\Box$ }

\newcommand{\lr}{\rightarrow}
\newcommand{\dsp}{\displaystyle}

\newcommand{\ess}{\mathrm{ess} \,}

\newcommand{\supp}{\operatorname{supp}}

\newcommand{\dive}{\operatorname{div}}

\newcommand{\eps}{\varepsilon}

\newcommand{\loc}{_{loc}}

\newcommand{\mN}{\mathbb{N}}

\newcommand{\mR}{\mathbb{R}}

\newcommand{\be}{{\bf e}}

%% file: No-lossy10-extension.bbl
\providecommand{\bysame}{\leavevmode\hbox to3em{\hrulefill}\thinspace}
\providecommand{\MR}{\relax\ifhmode\unskip\space\fi MR }
\providecommand{\MRhref}[2]{%
  \href{http://www.ams.org/mathscinet-getitem?mr=#1}{#2}
}
\providecommand{\href}[2]{#2}